\documentclass[english,aps,twocolumn,prd,superscriptaddress,showpacs,preprintnumbers,amsmath,amssymb]{revtex4}

\usepackage{subfigure}
\usepackage{amsmath,amssymb}
\usepackage{graphicx}
\usepackage{rotating}
\usepackage{dsfont}
\usepackage{color}
\usepackage[dvips]{epsfig}     
\usepackage{multirow}
\definecolor{refcol}{RGB}{0,0,205}
\usepackage{microtype}
\usepackage[colorlinks,linkcolor=refcol,citecolor=refcol,urlcolor=refcol]{hyperref}
\usepackage{breakurl}

\usepackage{charter} 
\usepackage[charter]{mathdesign}

\graphicspath{{./figures/}}

\def\di{\displaystyle}

\def\bg{\begin{eqnarray}\begin{array}{rcl}\displaystyle}
\def\eg{\end{array} &\di    &\di   \end{eqnarray}}
\def\bm#1{\begin{eqnarray}\begin{array}{#1}\di}
\def\bmo#1{\begin{eqnarray*}\begin{array}{#1}\di}
\def\bml#1#2{\begin{eqnarray}\begin{array}{#1}\label{#2}\di}
\def\bgo{\begin{eqnarray*}\begin{array}{rcl}\displaystyle}
\def\ego{\end{array} &\di    &\di \nonumber  \end{eqnarray*}}

\def\btensor#1#2{\renew\left#1\begin{array}{#2}\di}
\def\brtensor#1#2#3{\ren#3\left#1\begin{array}{#2}}
\def\botensor#1#2{\renew\left#1\begin{array}{#2}}
\def\etensor#1{\end{array}\right#1}

\def\eq#1{(\ref{#1})}

\def\Fig#1{Fig.~\ref{#1}}

\def\s0#1#2{\mbox{\small{$ \frac{#1}{#2} $}}}
\def\0#1#2{\frac{#1}{#2}}

\def\s{\sigma}

\def\ren#1{\renewcommand{\arraystretch}{#1}}

\def\renew{\renewcommand{\arraystretch}{1}}

\definecolor{blue}{rgb}{0,0,1}

\definecolor{green}{rgb}{0,1,0}

\definecolor{red}{rgb}{1,0,0}

\newcommand{\tr}{\mathrm{tr}}

\newcommand{\be}{\begin{eqnarray}}
\newcommand{\ee}{\end{eqnarray}}

\usepackage{babel}
\makeatother

\begin{document}

\title{The Vacuum Structure of Vector Mesons in QCD}
\author{Fabian Rennecke}
\email[E-mail: ]{f.rennecke@thphys.uni-heidelberg.de}
\affiliation{Institut f\"ur Theoretische Physik, University of Heidelberg, 
Philosophenweg 16, 62910 Heidelberg, Germany}
\affiliation{ExtreMe Matter Institute EMMI, GSI, Planckstr.~1, 64291 Darmstadt, Germany.}

\begin{abstract}
We study the chiral dynamics of vector mesons in two-flavor QCD in vacuum by utilizing a functional renormalization group approach. This allows us to capture the dynamical transition from the quark-gluon phase at high energies to the hadronic phase at low energies without the necessity of model parameter tuning. We use this to analyze the scaling of vector meson masses towards the chiral symmetry breaking scale, the decoupling of the mesons at high energies and the validity of vector meson dominance.
\end{abstract}

\maketitle

\section{Introduction}\label{sec:Intro}

The phase structure of Quantum Chromodynamics (QCD) is the subject of very active experimental and theoretical research. A crucial question is how to detect the formation of the quark-gluon plasma in heavy-ion collisions at ultrarelativistic energies. Vector mesons play a very important role in this context because they provide promising evidence for both deconfinement and chiral symmetry restoration. While the suppression of heavy quarkonium could be a signature of deconfinement \cite{Matsui:1986dk}, in-medium modifications of light vector mesons may signal chiral symmetry restoration \cite{Pisarski:1981mq}. The latter manifest themselves in low-mass dilepton data from heavy-ion collisions \cite{Adamova:2006nu,Arnaldi:2008fw,Geurts:2012rv}. Dileptons escape the fireball essentially without interaction and couple directly to light vector mesons such as the $\rho$. Thus, dilepton spectra show prominent vector meson peaks which allow for the investigation of in-medium modifications of these mesons \cite{Rapp:1999ej}. A connection between the modifications of vector mesons and chiral symmetry restoration in a hot and/or dense medium can be established e.g. by considering the scaling of the $\rho$ mass with temperature \cite{Brown:1991kk,Hatsuda:1991ez,Pisarski:1995xu} or the melting of the $\rho$ resonance \cite{Rapp:1997fs,Hohler:2013eba}. This connection is based on the fact that chiral symmetry restoration implies the degeneration of chiral partners such as $\rho$ and $a_1$. A thorough understanding of the dynamics of these mesons in QCD is therefore essential for a complete picture of the QCD phase structure.

In this work we present first results on the properties of the chiral partners $\rho$ and $a_1$ as they emerge from quark-gluon fluctuations at high energies. To this end, we study dynamical QCD based on \cite{Braun:2014ata}. It utilizes the functional renormalization group (FRG) approach to QCD \cite{Litim:1998nf,Berges:2000ew,Pawlowski:2005xe,Schaefer:2006sr,Gies:2006wv,Pawlowski:2010ht,Rosten:2010vm,Braun:2011pp,vonSmekal:2012vx,Pawlowski:2014aha}, extended by dynamical hadronization \cite{Gies:2001nw,Gies:2002hq,Floerchinger:2009uf,Pawlowski:2005xe}. This way, the properties of the hadrons are determined by the underlying dynamics of microscopic QCD and we can conveniently describe the transition from quarks and gluons to hadrons non-perturbatively without suffering from a fine-tuning of model parameters.

We concentrate on two-flavor QCD at vanishing temperature and density in Euclidean spacetime and develop a scale dependent effective action that captures the relevant dynamics in both, the quark-gluon phase and the hadron phase, on a qualitative level. Furthermore, we extend the dynamical hadronization technique to include vector mesons. The present analysis will serve as an starting point for qualitative and quantitative in-medium studies of vector mesons. It has been demonstrated in \cite{Mitter:2014wpa} for QCD in the quenched limit, that quantitative precision is indeed feasible with the FRG approach to QCD.

Even though we work in the vacuum, the renormalization group (RG) scale dependence of parameters such as the masses reflects their finite temperature behavior. In particular, there is a critical scale $k_\chi$ which separates the phases with broken and restored chiral symmetry. This allows us to study the behavior of hadronic parameters as they approach the scale of chiral symmetry restoration and clarify how mesons decouple from the physical spectrum at high energies. The scaling of the low-energy parameters is uniquely fixed from microscopic QCD. The reason is that the running of the hadronic parameters is governed by an infrared-attractive fixed point as long as the gauge coupling is small \cite{Gies:2002hq}. This guarantees that the memory of the initial conditions of the RG flows of these parameters, initiated at large, perturbatively accessible energy scales, is lost and the hadronic phase is uniquely determined within our truncation.

By exploiting this fact, we can analyze the validity of vector meson dominance (VMD) \cite{Sakurai:1960ju}. The idea of VMD is to promote the $SU(2)_L\!\times\! SU(2)_R$ flavor symmetry to a gauge symmetry. This way, $\rho$ and $a_1$ naturally appear as gauge bosons \cite{Lee:1967ug}. The main advantage is that VMD significantly reduces the number of different interactions involving vector mesons. The obvious conceptual shortcoming is that chiral symmetry is only a global symmetry in QCD. Furthermore, even though VMD has lead to accurate predictions in some cases at low energies \cite{Meissner:1987ge}, it gives e.g. the wrong phenomenology of $\rho$ and $a_1$ mesons \cite{Urban:2001ru}. We therefore keep chiral symmetry global and compare our results to the corresponding VMD predictions.

The paper is organized as follows: We motivate our ansatz for the quantum effective action used in this work in Sec.~\ref{sec: model}. In Sec.~\ref{sec:frg} we first introduce the FRG and dynamical hadronization in the presence of vector mesons. We continue with a discussion of the implications of $\pi\!-\!a_1$ mixing within our approach. At the end of this section, an outline of our treatment of the gauge sector of QCD is given. Our results are presented in Sec.~\ref{sec:results}. After discussing the initial conditions of the RG flows of our model, we present our results on the meson and quark masses with particular emphasis on the scaling of the $\rho$ mass towards the chiral symmetry breaking scale. Then, we show in more detail how the mesons disappear from the physical spectrum at large scales. Lastly, we discuss the validity of VMD. We end with a conclusion in Sec.~\ref{sec: concl} and provide some details about the RG flow equations in Sec.~\ref{sec:rucou}.

\section{The Scale Dependent Effective Action}\label{sec: model}

We aim at describing the chiral dynamics of two-flavor QCD at vanishing temperature and density and in particular the vacuum behavior of the vector meson chiral partners $\rho$ and $a_1$. Furthermore, we want to capture the dynamics of QCD over a wide range of scales, taking into account both quark-gluon dynamics in the quark-gluon phase at large energies and hadron dynamics in the low-energy hadronic regime. Thus, we base the construction of the effective action we use in this work on well-established renormalization group arguments. The starting point of our construction is the microscopic gauge fixed action of QCD. Owing to the increasing strength of the strong coupling towards lower scales, effective four quark interactions are dynamically generated and become physically relevant. They drive the chiral phase transition, as poles in these quark-antiquark scattering channels signal the formation of bound states and the breaking of chiral symmetry.

The tensor structure of the four-quark interaction channels is directly related to the quantum numbers of the corresponding mesons that are formed in the respective scattering channel. Therefore, we need to include at least those interaction channels, that carry the quantum numbers of the mesons we are interested in. In the present case, these are the Lorentz--scalar-pseudoscalar iso--scalar-vecor and the Lorentz--vector-axialvector iso--vector-vector channels $\lambda_{S,k}$ and $\lambda_{V,k}$. On the mesonic side, these channels correspond to the sigma and the pion, and the rho and the $a_1$ respectively. 

The scalar-pseudoscalar channel is the dominant channel among all possible quark-antiquark scattering channels in vacuum. This has been explicitly checked by considering a complete basis of four-quark interactions \cite{Mitter:2014wpa}. This implies in particular, that the pions and the sigma mesons dominate the dynamics in the hadronic phase. In this work we demonstrate explicitly on the example of vector mesons, that there is an emergent scale hierarchy where only the lightest mesons, i.e. pions and sigma, can contribute to the dynamics of the system at low energies. Thus, the properties of the heavier meson states in Euclidean space are completely fixed by quark-gluon dynamics at large energies and pion-sigma dynamics at low energy scales. 

To properly take into account the dynamics in the hadronic phase, we model this sector by an effective meson potential which in principle includes arbitrary orders of mesonic self-interactions. Furthermore, we consider momentum dependent propagators of the quarks and mesons, based on a small-momentum expansion, by including scale dependent wave function renormalizations $Z_k$. To account for non-vanishing current quark masses, a source term $-c\sigma$ in the meson sector explicitly breaks chiral symmetry. It is directly related to finite current quark masses. As a consequence, pions are massive rather than Goldstone bosons and the chiral transition is a crossover.

To connect the the quark sector with the meson sector, we include scalar channel and vector channel Yukawa couplings $h_{S,k}$ and $h_{V,k}$. In order to consistently account for the dynamical change of degrees of freedom from the quark-gluon phase to the hadronic phase, we use dynamical hadronization as it was put forward in \cite{Braun:2014ata}. We will elaborate on this in the next section. As we will demonstrate there, it is inevitable to use this formulation here, since the the elimination of the $\pi\!-\!a_1$ mixing results in manifestly scale dependent $a_1$ fields.

In summary, we use the following scale dependent effective action:
\begin{widetext}
\begin{align}\label{eq:trunc}
 \Gamma_k =& \int_x\biggl\{Z_{\psi,k}\, \bar{q} \left(i \gamma_\mu D_\mu\right) q+ \frac{1}{4}F^a_{\mu \nu} F^a_{\mu \nu} + \bar c^a \partial_\mu D^{ab}_\mu c^b + \frac{1}{2\xi} (\partial_\mu A_\mu^a)^2+\Delta \mathcal{L}_{\text{glue}}\biggr. \\ \nonumber
  & + \biggl. \frac{\lambda_{S,k}}{2} \left[(\bar{q} q)^2-(\bar{q} \gamma_5 \vec{\tau} q)^2\right]\!-\frac{\lambda_{V,k}}{2} \left[(\bar{q} \gamma_\mu\vec{\tau} q)^2+(\bar{q} \gamma_\mu\gamma_5 \vec{\tau} q)^2\right]\!+\biggl. h_{S,k} \left[ \bar{q} (\gamma_5\vec{\tau}\vec{\pi}+i\sigma) q \right]\! + h_{V,k} \left[ \bar{q} (\gamma_\mu\vec{\tau}\,\vec{\rho}^{\,\mu}+\gamma_\mu\gamma_5\vec{\tau}\,\vec{a}_1^{\,\mu}) q \right] \biggr.\\ \nonumber
& +\biggl.\frac{1}{2} Z_{S,k} (\partial_{\mu} \varphi)^2+\frac{1}{8} Z_{V,k}\, \tr\left(\partial_\mu V_\nu-\partial_\nu V_\mu\right)^2 + U_k(\varphi,V_\mu) \biggr\}\,,
\end{align}
\end{widetext}
with the covariant derivative $D_\mu = \partial_\mu - i Z_{A,k}^{1/2}g_k\, A_\mu^a t^a$, where $g_k = \sqrt{4 \pi \alpha_{s,k}}$ is the strong coupling and $Z_{A,k}$ the gluonic wave function renormalization. With this definition the covariant derivative $D_\mu$ is renormalization group invariant. We use Hermitian gamma matrices so that $\{\gamma_{\mu},\gamma_{\nu}\}=2 \delta_{\mu \nu} \mathds{1}$. The commutator for the $SU(N_c)$ generators reads $[t^a,t^b]=i f^{abc} t^c$ and, hence, the trace is positive, $\text{Tr}\, t^a t^b=\frac{1}{2} \delta^{ab}$. $\vec{\tau}$ are the Pauli matrices. For the field strength tensor we use the relation
\begin{align}
F_{\mu\nu} = \0{i}{Z^{1/2}_{A,k} g_{k}  } [ D_{\mu}\,,\,D_{\nu}]\,.
\end{align}
We work in Landau gauge here, so $\xi = 0$. For more details on the gauge part of our truncation see Sec.~\ref{sec:gauge}.

The first line of \eq{eq:trunc} contains the microscopic gauge fixed action of QCD. As mentioned above, we introduced a running quark wave function renormalization $Z_{\psi,k}$ to capture some non-trivial momentum dependence of the quark propagator. $\Delta\mathcal{L}_{\text{glue}}$ stands for the fluctuation-induced part of the full momentum dependence of ghost and gluon propagators as well as non-trivial ghost-gluon, three-gluon and four-gluon vertex corrections, for details see \cite{Fister:2013bh,Fister:2011uw,Fister:Diss} and Sec.~\ref{sec:gauge}.

The four-quark interaction channels and the corresponding Yukawa interactions are in the second line of \eq{eq:trunc}. The Yukawa sector arises from the bosonization of the quark sector. With dynamical hadronization as explained in the next section, these interactions will basically carry the quark self-interactions in the quark-gluon regime (see also \eq{eq:modhs} and \eq{eq:modhv}).

The third line of \eq{eq:trunc} contains the meson sector of our truncation. With the running wave function renormalizations $Z_{S,k}$ and $Z_{V,k}$ for the scalar and vector mesons respectively, we capture the major part of the momentum dependence of the full meson propagators \cite{Helmboldt:2014iya}. Furthermore, as we explicitly demonstrate in Sec.~\ref{sec:mesdec}, the wave function renormalizations play a crucial role for the decoupling of the mesons at high energies. They are therefore indispensable for the identification of the physical meson masses.

The meson interactions are stored in $U_k(\varphi,V_\mu)$, which reads
\begin{align}\label{eq:pot}
\begin{split}
&U_k(\varphi,V_\mu) =\\
 &\quad \frac{1}{2}m_{S,k}^2\left(\varphi^2-\varphi_0^2\right)+\frac{1}{8}\nu_k\left(\varphi^2-\varphi_0^2\right)^2-c\sigma\\
&\quad - i g_{1,k} V_\mu\varphi \!\cdot\! \partial_\mu\varphi -\frac{1}{2}g_{2,k} \left(V_\mu\varphi\right)^2 + \frac{1}{4}g_{3,k}\varphi^2\tr\,V_\mu V_\mu\\
&\quad+\frac{1}{4}m_{V,k}^2 \tr\, V_\mu V_\mu - \frac{i}{2}g_{4,k}\,\tr\, \partial_\mu V_\nu [V_\mu,V_\nu]\\
&\quad - \frac{1}{4} g_{5,k}\, \tr\, V_\mu V_\nu [V_\mu,V_\nu]\,.
\end{split}
\end{align}
Since within this work our focus is on the qualitative physical picture of vector mesons in vacuum, we restrict $U_k$ to relevant and marginal interactions along the lines of \cite{Urban:2001ru}. The effect of irrelevant operators, which are potentially non-negligible if one is interested in quantitative precision \cite{Pawlowski:2014zaa}, is considered elsewhere. Note that $U_k$ is not an effective potential, since it also contains the term $\sim g_{1,k}$ with an explicit derivative of the scalar meson field.

We use an $O(4)$ representation for the meson fields:
\begin{align}
\varphi=\begin{pmatrix}\vec{\pi}\\ \sigma\end{pmatrix},\quad V_\mu = \vec{\rho}^{\,\mu}\vec{T}+\vec{a}_1^{\,\mu}\vec{T}^5,
\end{align}
and the vacuum expectation value of the mesons is given by $\varphi_0 = (0,\sigma_{0,k})^T$. We define the $\mathfrak{so}(4)$ matrices
\begin{align}
(T_i)_{jk}=\begin{pmatrix} -i\epsilon_{ijk} & \vec{0} \\ \vec{0}^T & 0\end{pmatrix},\quad (T_i^5)=\begin{pmatrix} 0_{3 \!\times\! 3} & -i\vec{e}_i \\ i\vec{e}_i^{\,T} & 0\end{pmatrix},
\end{align}
with $i,j,k \in \{1,2,3\}$ and $\vec{e}_i^{\,T} = (\delta_{1i},\delta_{2i},\delta_{3i})$. They obey the following commutation relations:
\begin{align}
\begin{split}
[T_i,T_j]&=i \epsilon_{ijk} T_k,\\
[T_i^5,T_j^5]&=i \epsilon_{ijk} T_k,\\
[T_i,T_j^5]&=i \epsilon_{ijk} T_k^5,
\end{split}
\end{align}
and therefore $T_i^L=\frac{1}{2}(T_i-T_i^5)$ and $T_i^R=\frac{1}{2}(T_i+T_i^5)$ form representations of $SU(2)_L$ and $SU(2)_R$. The mesons transform under these flavor rotations as
\begin{align}
\varphi \rightarrow U\varphi\,,\quad V_\mu \rightarrow U V_\mu U^\dagger\,,
\end{align}
with $U=e^{i\vec{\alpha}\vec{T}+i\vec{\beta}\vec{T}^b}$, where $\vec{\alpha}$ and $\vec{\beta}$ are arbitrary vectors. For more details on the meson sector of our truncation, we refer to \cite{Urban:2001ru}.

With this definitions the scalar-vector and vector-vector meson interactions in \eq{eq:pot} can be written as follows:
\begin{align}\label{eq:vints}
\begin{split}
- i g_{1,k} V_\mu\varphi\cdot\partial_\mu\varphi &= g_{1,k}\Bigl[(\vec{\rho}^{\,\mu}\!\times\!\vec{\pi})\cdot\partial_\mu\vec{\pi}-\sigma\vec{a}_1^{\,\mu}\cdot\partial_\mu\vec{\pi}\Bigr.\\
&\quad+\Bigl.\vec{a}_1^{\,\mu}\cdot\vec{\pi}\partial_\mu\sigma  \Bigr]\,,\\
-\frac{g_{2,k}}{2} \left(\!V_\mu\varphi\!\right)^2\! &= \frac{g_{2,k}}{2}\Bigl[\!  (\vec{\rho}^{\,\mu}\!\times\!\vec{\pi}\!-\!\sigma\vec{a}_1^{\,\mu})^2\!+(\vec{a}_1^{\,\mu}\!\cdot\!\vec{\pi})^2\! \Bigr]\,,\\
\frac{g_{3,k}}{4}\varphi^2\tr\,V_\mu V_\mu &= \frac{g_{3,k}}{2}\left(\vec{\pi}^{\, 2}+\sigma^2\right)\left((\vec{\rho}^{\,\mu})^2+(\vec{a}_1^{\,\mu})^2\right)\,,\\
- \frac{i}{2}g_{4,k}\,\tr\, \partial_\mu V_\nu [V_\mu,V_\nu] &= g_{4,k}\,\bigl[ \partial_\mu\vec{\rho}^{\nu}\!\cdot\! \left( \vec{\rho}^{\,\mu}\!\times\!\vec{\rho}^{\nu} + \vec{a}_1^{\,\mu}\!\times\!\vec{a}_1^{\nu} \right)\bigr.\\
&\quad +\bigl. \partial_\mu\vec{a}_1^{\nu}\!\cdot\! \left( \vec{\rho}^{\,\mu}\!\times\!\vec{a}_1^{\nu} - \vec{a}_1^{\,\mu}\!\times\!\vec{\rho}^{\nu} \right) \bigr]\,,\\
- \frac{1}{4} g_{5,k}\, \tr\, V_\mu V_\nu [V_\mu,V_\nu] &= \frac{g_{5,k}}{4}\left[ \left( \vec{\rho}^{\,\mu}\!\times\!\vec{\rho}^{\nu} + \vec{a}_1^{\,\mu}\!\times\!\vec{a}_1^{\nu} \right)^2\right.\\
&\quad +\left. \left( \vec{\rho}^{\,\mu}\!\times\!\vec{a}_1^{\nu} - \vec{a}_1^{\,\mu}\!\times\!\vec{\rho}^{\nu} \right)^2 \right]\,.
\end{split}
\end{align}

As it is evident from \eq{eq:pot} and mentioned in the introduction, we do not assume VMD. This gauge principle would lead to to the following relations between the different couplings of our truncation:
\begin{align}\label{eq:vmd}
g_{1,k}^2=g_{2,k}=g_{4,k}^2=g_{5,k} \quad \text{and} \quad g_{3,k}=0.
\end{align}
By inspection of the renormalization group flow of these couplings, we will show that VMD would lead to an oversimplification of the dynamics of the system. Nonetheless, VMD turns out to be a good approximation at low energies, see Sec~\ref{sec:vmd}.

In the present setup, the masses of the quarks and the mesons are given by
\begin{align}\label{masses}
\begin{split}
m_{q,k}^2 &= h_{s,k}^2 \sigma_{0,k}^2\,,\\
m_{\pi,k}^2 &= m_{S,k}^2\,,\\
m_{\sigma,k}^2 &= m_{S,k}^2+\lambda_{4,k} \sigma_{0,k}^2\,,\\
m_{\rho,k}^2 &= m_{V,k}^2+g_{3,k}\sigma_{0,k}^2\,,\\
m_{a_1,k}^2 &= m_{V,k}^2+(g_{2,k}+g_{3,k})\sigma_{0,k}^2\,.
\end{split}
\end{align}
We see that the $\pi$ and the $\sigma$ meson as well as the $\rho$ and $a_1$ meson have degenerate masses in the chirally symmetric phase which is characterized by $\sigma_{0,k} = 0$. When chiral symmetry is broken, this degeneracy is lifted. The mass-splitting of the scalar mesons is then determined by the quartic scalar meson coupling $\lambda_{4,k}$. The mass-splitting of the vector mesons is determined by the strength of the interaction $g_{2,k}$. Note that, owing to the symmetry breaking source $c>0$, we are not in the chiral limit. Thus, the chiral order parameter $\sigma_{0,k}$ is always nonzero.

Even though the masses we extract here are the curvature masses, it was shown in \cite{Helmboldt:2014iya} on the example of the pion mass in a quark meson model, that the curvature mass of the mesons is almost identical to the pole mass for truncations that include running wave function renormalizations. Thus, as mentioned above, we capture the major part of the momentum dependence of the full meson propagators by including $Z_{S,k}$ and $Z_{V,k}$ and the masses are very close to the physical masses.

We note that even though the action contains massive vector bosons, it is not necessary to use the Stueckelberg formalism to ensure renormalizability \cite{itzykson2006quantum}. UV regularity is always guaranteed for the functional renormalization group, as long as the scale derivative of the regulator decays fast enough for momenta much larger than the cutoff scale.

\section{Fluctuations and the transition from quarks to mesons}\label{sec:frg}

In this work we are interested in the dynamical transition from UV to IR degrees of freedom. To achieve this, we include quantum fluctuations by means of the functional renormalization group. For QCD related reviews see \cite{Litim:1998nf,Pawlowski:2005xe,Gies:2006wv,Pawlowski:2010ht,
Rosten:2010vm,Berges:2000ew,Schaefer:2006sr,Braun:2011pp,vonSmekal:2012vx}. Furthermore, in order to consistently describe the dynamical change of degrees of freedom, we use dynamical hadronization \cite{Braun:2014ata,Gies:2001nw,Gies:2002hq,Floerchinger:2009uf,Pawlowski:2005xe}. This allows for a unified description of the interplay between different degrees of freedom at different scales in terms of a single effective action.

\subsection{Functional renormalization group and dynamical hadronization in the presence of vector mesons}

Here, we follow the discussion given in \cite{Braun:2014ata}. In addition, since this work constitutes the first FRG study of vector mesons in QCD, we will discuss the implication for the flow equations and dynamical hadronization in this case.

The starting point of the functional renormalization group is the
scale-dependent effective action $\Gamma_\Lambda$ at a UV-cutoff scale
$\Lambda$. In the case of first-principle QCD, $\Lambda$ is a large,
perturbative energy scale and correspondingly $\Gamma_\Lambda$ is the
microscopic QCD action with the strong coupling constant and the
current quark masses as the only free parameters. Quantum
fluctuations are successively included by integrating out momentum
shells down to the RG-scale $k$. This yields the scale-dependent
effective action $\Gamma_k$, which includes fluctuations from momentum
modes with momenta larger than $k$. By lowering $k$ we resolve the
macroscopic properties of the system and eventually arrive at the full
quantum effective action $\Gamma = \Gamma_{k=0}$. The RG-evolution of
the scale-dependent effective action is given by the Wetterich
equation \cite{Wetterich:1992yh}.

As we have discussed above, a formulation of the effective action in terms of local composite fields is more efficient in the hadronic phase of QCD. In order to dynamically connect this regime with the ultraviolet regime of QCD, where quarks and gluons are the dynamical fields, we use dynamical hadronization as it was put forward in \cite{Braun:2014ata}. This implies that the meson fields in \eq{eq:trunc} are RG-scale dependent. This yields a modified Wetterich equation, which reads with $\Phi=(A,q,\bar{q},c,\bar c,\pi,\sigma,\rho,a_1)$ in a shorthand notation:
\begin{align}\label{eq:fleq}
\partial_t\Gamma_k[\Phi]=\frac{1}{2}\text{Tr}\left[ \left( \Gamma_k^{(2)}[\Phi]+R_k^{\Phi} \right)^{-1}\!\cdot\partial_t R_k^{\Phi}\right] - \frac{\delta \Gamma_k}{\delta\phi_i}\cdot\partial_t\phi_i\,,
\end{align}
where $\phi=(\pi,\sigma,\rho,a_1)$ summarizes the meson fields. $\partial_t$ is the total derivative with respect to the RG-time
$t=\ln( k/\Lambda)$ and the traces sum over discrete and continuous
indices of the fields, including momenta and species of
fields. This also includes the characteristic minus sign and a factor of 2 for fermions. $\Gamma_k^{(2)}[\Phi]$ denotes the second functional derivative of the effective action with respect to all combinations of the fields. $R_k^\Phi$ is the regulator function for the field $\Phi$. It is diagonal in field space. Note that in order not to break chiral symmetry explicitly by our regularization scheme, we introduced the same regulators for the scalar mesons and the vector mesons respectively. For details we refer to App.~\ref{sec:rucou}. The flow equation can be written schematically as
\begin{align}\label{fig:fleq}
\includegraphics[height=8.7ex]{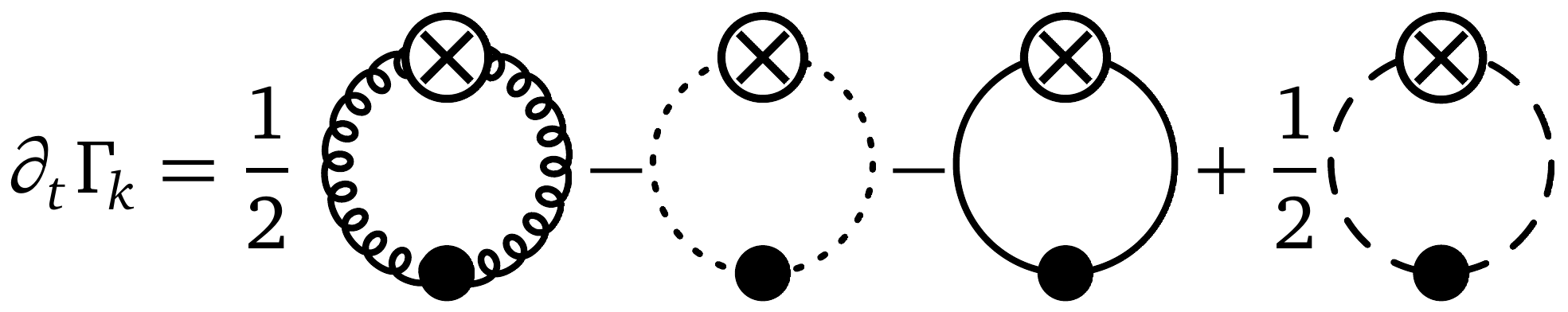}
\end{align}
Here, the first three diagrams represent the gluon, ghost and quark contributions to the flow equation and the fourth diagram depicts the meson contributions. The black dot indicates that the propagators are the full propagators. The crossed circles represent the regulator insertion. By inserting the truncation \eq{eq:trunc} into the flow equation \eq{eq:fleq}, one gets a closed set of fully coupled RG-flow equations for the scale-dependent parameters of the truncation.

$\frac{\delta \Gamma_k}{\delta\phi_i}\cdot\partial_t\phi_i$ stands for the modifications of the flow equation due to dynamical hadronization. The general idea is to store four-quark interactions entirely in the meson sector at every scale $k$, i.e. we perform a bosonization at every scale \cite{Gies:2001nw,Gies:2002hq,Floerchinger:2009uf,Pawlowski:2005xe}. The hadronization fields $\phi$, which carry the quantum numbers of the mesons, become scale dependent and can be viewed as hybrid fields: while they encode the quark dynamics at large energies, they behave as mesons at low energies. This way, a dynamical connection between these two distinct sectors is established and the transition scale is an emergent scale which is fixed by the fluctuations of fundamental QCD. This is in contrast to conventional bosonization by means of a Hubbard-Stratonovich transformation, where four-quark interactions are stored in the meson sector at a fixed scale and therefore four quark-interactions are neglected/mis-counted at other scales \cite{Gies:2006wv}.

The gauge sector as well as the quarks are not affected by the hadronization. The scale dependence of the mesons is given by their flow $\partial_t\phi_i$, which reads for the individual mesons
\begin{align}\label{eq:hadflow}
\begin{split}
\partial_t\vec{\pi} &= \dot A_k \bar{q}\gamma_5\vec{\tau}q\,,\\
\partial_t\sigma &= \dot A_k \bar{q}i q\,,\\
\partial_t\vec{\rho}^{\,\mu} &= \dot B_k \bar{q}\gamma_\mu\vec{\tau}q\,,\\
\partial_t\vec{a}_1^{\,\mu} &= \dot B_k \bar{q}\gamma_\mu\gamma_5\vec{\tau}q - \dot C_k\partial_\mu\vec{\pi}\,.
\end{split}
\end{align}
Note that the structure of the mesons as quark-antiquark bilinears becomes apparent in this formulation. $\dot A_k$ and $\dot B_k$ are the hadronization functions. Their precise form is not determined a priori. In accordance with the discussion above, we fix them such that the fermionic self-interactions that drive chiral symmetry breaking and reflect the meson content of our theory are stored in the meson sector. Thus, the the four-quark interactions are completely absorbed into the meson sector, enforcing
\begin{align}\label{eq:hadcond} 
\partial_t \lambda_{S,k}=0\quad \text{and}\quad \partial_t \lambda_{V,k}=0\,.
\end{align}
Note that this formulation eliminates all double- and/or mis-counting problems, which potentially occur in models including both quark and hadron degrees of freedom \cite{Braun:2014ata}. \eq{eq:hadcond} yields the following hadronization functions:
\begin{align}\label{eq:hadfunc}
\begin{split}
\dot A_k &= -\frac{1}{2 h_{S,k}}\bigl.\partial_t\bigr|_{\phi}\lambda_{S,k}\,\\
\dot B_k &= -\frac{1}{2 h_{V,k}}\bigl.\partial_t\bigr|_{\phi}\lambda_{V,k}\,,
\end{split}
\end{align}
where $\bigl.\partial_t\bigr|_{\phi}$ denotes the scale derivative with fixed hadronization fields. These hadronization functions give rise to modified running couplings of \eq{eq:trunc}. They are given in App.~\ref{sec:rucou}.

In addition to the quark-bilinear term with the quantum numbers of the corresponding meson in \eq{eq:hadflow}, the flow of the $a_1$-meson has an additional contribution proportional to $\partial_\mu\vec{\pi}$. This term arises because the so-called $\pi\!-\!a_1$ mixing leads to an additional scale dependence of the $a_1$ meson, which has to be taken into account and fixes $\dot C_k$. We will elaborate on this point in the next section.

\subsection{$\pi\!-\!a_1$ mixing}\label{sec:pia1}

Spontaneous chiral symmetry breaking leads to a non-vanishing vacuum expectation value $\sigma_0$ of the $\sigma$ meson and the resulting mixing term
\begin{align}\label{eq:pia1}
\Gamma_{\pi a_1}=-\int_x g_{1,k}\, \sigma_0\, \vec{a}_1^{\,\mu}\!\cdot\!\partial_\mu\vec{\pi}\,,
\end{align}
implies an off-diagonal two-point function $\Gamma_k^{(2)}$. This is referred to as $\pi\!-\!a_1$ mixing. Here, we will eliminate this mixing by a redefinition of the $a_1$ field,
\begin{align}\label{eq:pia1rep}
\vec{a}_1^{\,\mu} \longrightarrow \vec{a}_1^{\,\mu}+\frac{g_{1,k} \sigma}{m_{V,k}^2+(g_{2,k}+g_{3,k})\sigma^2}\, \partial_\mu\vec{\pi}\,.
\end{align}
This redefinition of the $a_1$ field renders it explicitly RG-scale dependent, $\partial_t \vec{a}_1^{\,\mu}\neq 0$. Before we discuss the implications of this scale dependence, we turn toward the resulting modifications of the effective action \eq{eq:trunc}. If we plug \eq{eq:pia1rep} into the truncation \eq{eq:trunc}, the part of the action leading to the mixing term \eq{eq:pia1} is canceled and various new terms appear. Since the replacement \eq{eq:pia1rep} introduces terms $\sim \partial_\mu \vec{\pi}$, the interactions of our original ansatz \eq{eq:trunc} receive modifications with explicit momentum dependence. Within this work, we define all running coupling at vanishing external momentum, see App.~\ref{sec:rucou}. Thus, for interactions that are not explicitly momentum dependent in our original action, these modifications simply drop out of the beta functions. Only the meson anomalous dimension $Z_{S,k}$ and the scalar-scalar vector interaction $g_{1,k}$ receive non-vanishing modifications. The new term $\sim (\partial_\mu \vec{\pi})^2$ yields for the pion wave function renormalization
\begin{align}\label{eq:zpim}
Z_{\pi,k} = Z_{S,k} - \frac{g_{1,k}^2 \sigma_{0,k}^2}{m_{a_1}^2}\,.
\end{align}
While the wave function renormalizations do not enter RG-invariant beta functions, their the anomalous dimension $\eta_k = -(\partial_t Z_k)/Z_k$ does. Thus, \eq{eq:zpim} yields a modified pion anomalous dimension.

The other relevant modification affects the $\rho\pi\pi$ vertex, which now reads
\begin{align}\label{eq:rppm}
\Gamma_{\rho\pi\pi}^{(3)}= g_{1,k} \left( 1-\frac{g_{2,k} \sigma_{0,k}^2}{m_{a_1,k}^2} \right)\,.
\end{align}
Since we define the coupling $g_{1,k}$ via this vertex, this has to be taken into account in the corresponding beta function, see App.~\ref{sec:rucou}.

The elimination of the $\pi\!-\!a_1$ mixing entails a shift of the $a_1$ field which includes running couplings. As a consequence, the $a_1$ field becomes RG-scale dependent itself. As we have discussed in the previous section, we use the dynamical hadronization technique which implies that all meson fields are scale dependent. The scale dependence of $a_1$ induced by \eq{eq:pia1rep} is additional to the one induced by dynamical hadronization. The total scale dependence of $a_1$ is now given by the RG flow
\begin{align}\label{eq:dta1}
\partial_t \vec{a}_1^{\,\mu} = \dot{B}_k \bar{q} \gamma_\mu\gamma_5\vec{\tau}q - \dot{C}_k \partial_\mu \vec{\pi}\,.
\end{align}
The first term stems from dynamical hadronization and reflects the quark-bilinear nature of the $a_1$ meson. The hadronization function $\dot{B}_k$ is given in Eq.~\eq{eq:hadfunc}. The second term is a result of the diagonalization of the meson two-point function and according to \eq{eq:pia1rep} $\dot{C}_k$ is given by
\begin{align}\label{eq:cdot}
\dot{C}_k = \partial_t\left( \frac{g_{1,k}\, \sigma}{m_{V}^2+(g_{2,k}+g_{3,k})\sigma^2} \right)\,.
\end{align}
In summary, chiral symmetry breaking leads to an off-diagonal meson two-point function. Diagonalization leads to modifications of the pion anomalous dimension and the $\rho\pi\pi$ interaction and introduces an additional scale dependence to the $a_1$ meson.

\subsection{Gauge sector}\label{sec:gauge}

Here, we briefly discuss the gauge sector of our truncation \eq{eq:trunc}. We follow the approach to dynamical QCD which was put forward in \cite{Braun:2014ata} and refer to this work for further details.

The gauge couplings induced by three-point functions play a dominant role in the description of interactions. Most importantly, effective four-quark interactions are generated by strong quark-gluon interactions during the RG flow in the quark-gluon regime. As discussed above, these four-quark interactions drive the chiral phase transition and the formation of mesons in the corresponding quark-antiquark scattering channels. We therefore solve the full flow equations for all three-point functions in QCD, i.e. the quark-gluon vertex $g_{\bar q A q,k}$, the three-gluon vertex $g_{A^3,k}$ and the ghost gluon vertex $g_{\bar c A c,k}$.

In order to keep the description as simple as possible, despite the rather large number of running couplings we consider here, we restrict our analysis to vanishing external momentum in the $n$-point functions. Together with an appropriate choice of the regulator function $R_k$, this yields analytical flow equations, see App.~\ref{sec:rucou}. Furthermore, only the classical tensor structures are taken into account in the gauge sector. It has been shown in \cite{Mitter:2014wpa} for quenched QCD, that both non-trivial momentum dependencies and non-classical tensor structures in the gauge sector lead to important quantitative corrections resulting in larger gauge couplings in favor of stronger chiral symmetry breaking. To phenomenologically account for these effects, we introduce an IR-strength function $\varsigma(k)$ which smoothly increases the gauge couplings by a factor $\sim 1.17$ in the non-perturbative regime at scales $k<2\,\text{GeV}$, while it leaves the gauge couplings unchanged in the perturbative regime. For more details we refer to \cite{Braun:2014ata}.

\begin{figure}[t]
\begin{center}
  \includegraphics[width=.98\columnwidth]{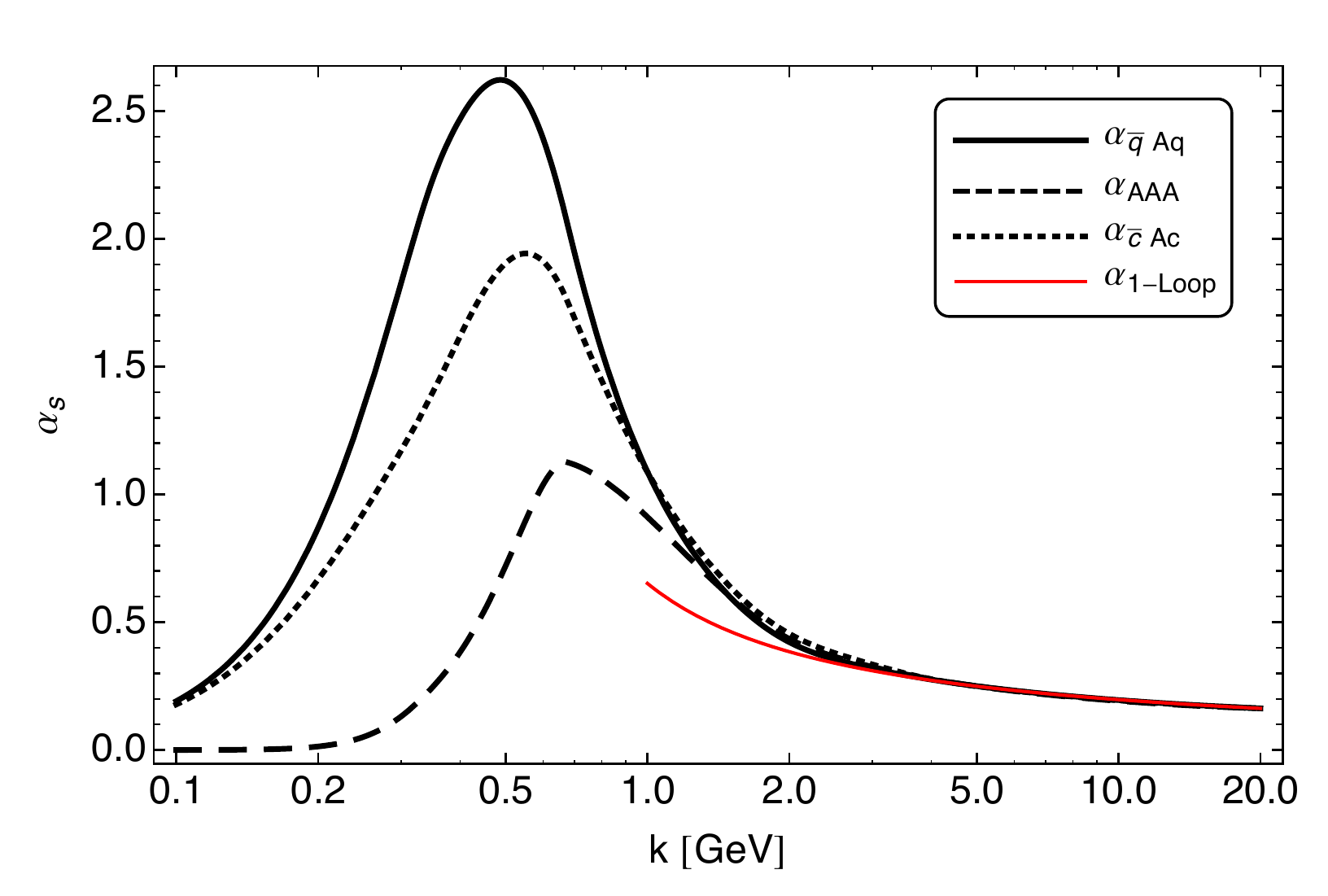}
  \caption{The running of the different strong couplings in comparison
    to the 1-loop running.}\label{fig:alphas}
\end{center}
\end{figure}

We also include four-gluon interactions via the four gluon vertex $g_{A^4,k}$. For the sake of simplicity, we choose a semi-perturbative ansatz that ensures the correct perturbative running and set
\begin{align}\label{eq:g4a}
g_{A^4,k}=g_{A^3,k}\,.
\end{align}
This approximation is valid for semi-perturbative scales and larger, $k\gtrsim 1.5~\text{GeV}$. At smaller scales, non-perturbative effects potentially lead to a different running.

In summary, we consider the following vertices in the gauge sector:
\begin{align}\label{eq:vertices}
\begin{split}
  \Gamma_k^{(\bar{q} A q)}&=Z_{A,k}^\frac{1}{2} Z_{q,k}\, g_{\bar{q} A q,k}\ S^{(3)}_{\bar{q} A q}\,,\\
  \Gamma_k^{(A^3)}&=Z_{A,k}^\frac{3}{2}\, g_{A^3,k}\ S^{(3)}_{A^3}\,,\\
  \Gamma_k^{(A^4)}&=Z_{A,k}^2\, g^2_{A^3,k}\ S^{(4)}_{A^4}\,,\\
  \Gamma_k^{(\bar{c} A c)}&=Z_{A,k}^\frac{1}{2} Z_{c,k}\, g_{\bar{c} A
    c,k}\ S^{(3)}_{\bar{c} A c}\,.
\end{split}
\end{align}
The classical tensor structures $S^{(n)}_{\Phi_1...\Phi_n}$ are obtained from the gauge fixed classical QCD action $\Gamma_\Lambda$ by
\begin{align}
S^{(n)}_{\Phi_1...\Phi_n} = \left. \frac{\delta^{n} \Gamma_{\Lambda}}{{\delta\Phi_1 \ldots \delta \Phi_n}} \right|_{g_k=1}\,,
\end{align}
where we have omitted indices for clarity. The definitions of the flow equations of the gauge couplings can be found in App.~\ref{sec:rucou}. The different running gauge couplings follow the perturbative running for scales $k\gtrsim 4\,\text{GeV}$, while non-perturbative effects lead to a different running at smaller scales. \Fig{fig:alphas} shows our results for the running gauge couplings.

For the gluon and ghost propagators we use the two-point functions
\begin{align}
\Gamma_{A/c,k}^{(2),{\rm YM}}(p) = \frac{1}{Z^{\text{YM}}_{A/c,k}(p^2) p^2+R_k^{A/c}}\,,
\end{align}
computed in \cite{Fischer:2008yv,FP} for pure YM theory as input, and include matter back-reactions in order to describe unquenched QCD \cite{Braun:2014ata}. We make use of the fact that within our construction the propagators enter the flow equation only via the corresponding anomalous dimensions
\begin{align}
\eta_{A/c,k} = -\frac{\partial_t Z_{A/c,k}}{Z_{A/c,k}}\,,
\end{align}
where we only consider external momenta $p\!=\!k$ and define $Z_{A/c,k}=Z_{A/c,k}(k^2)$. The anomalous dimensions of YM are functions of the YM gauge couplings. Thus, in our case, where we identified the four-gluon running coupling with the three-gluon running coupling, we have
\begin{align}\label{eq:etaYM}
\eta_{A/c,k}^{\text{YM}}=\eta_{A/c,k}^{\text{YM}}\left( g_{\bar c A c,k}^\text{YM}, g_{A^3,k}^\text{YM} \right)\,.
\end{align} 
The gluon anomalous dimensions $\eta_{A,k}$ of full QCD consists of a pure gauge part $\eta_{\text{glue},k}$ and the vacuum polarization $\Delta\eta_{A,k}$ induced by quark fluctuations,
\begin{align}
\eta_{A,k} = \eta_{\text{glue},k} + \Delta\eta_{A,k}\,.
\end{align}
We include the full vacuum polarization at vanishing external momentum. For details see App.~\ref{sec:rucou}. For the pure glue contribution of dynamical QCD we make use of \eq{eq:etaYM}: Since we know that the anomalous dimensions can be expressed as functions of the gauge couplings, we replace the the pure YM gauge couplings in \eq{eq:etaYM} by those of dynamical QCD,
\begin{align}
\eta_{\text{glue},k} = \eta_{A,k}^{\text{YM}}\left( g_{\bar c A c,k}, g_{A^3,k}\right)\,.
\end{align}
In practice, we use the approximation $g_{\bar c A c,k}^\text{YM} = g_{A^3,k}^\text{YM}$, which is valid down to the semi-perturbative regime, and parametrize $\eta_{A,k}^{\text{YM}}$ by only one coupling, i.e. $\eta_{A,k}^{\text{YM}}\approx\eta_{A,k}^{\text{YM}}(g_{\bar c A c,k}^\text{YM})$. Then, we can use a numerical fit function $\eta_{A,k}^{\text{YM}}(g)$ extracted from the input \eq{eq:etaYM} and insert the corresponding QCD coupling. This yields our final expression for the gluon anomalous dimension of QCD:
\begin{align}
\eta_{A,k} = \eta_{A,k}^{\text{YM}}(g_{\bar c A c,k}) + \Delta\eta_{A,k}\,.
\end{align}

We proceed analogously for the ghost anomalous dimension. There, no direct quark corrections arise in QCD and we only need to adapt the input anomalous dimension from YM to QCD. As it turns out, $\eta_{c,k}^{YM}$ is very well described by a function linear in $g_k^2$. We therefore find for the ghost anomalous dimension of dynamical QCD:
\begin{align}
\eta_{c,k} = \left(\frac{g_{\bar c A c,k}}{g_{\bar c A c,k}^\text{YM}}\right)^2 \eta_{c,k}^\text{YM}.
\end{align}

\section{Results}\label{sec:results}

\subsection{Initial Conditions}

\begin{figure}[t]
\begin{center}
  \includegraphics[width=0.98\columnwidth]{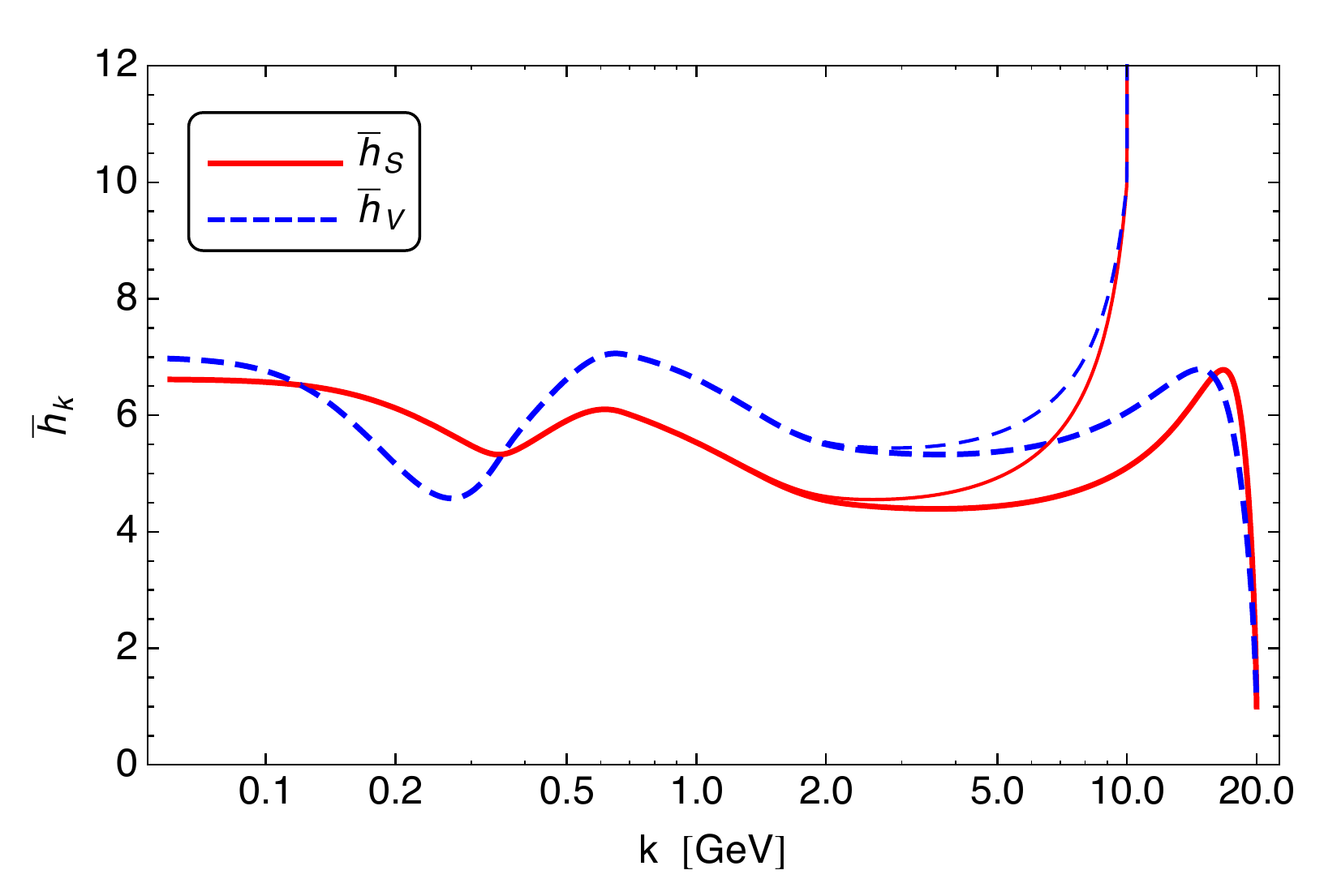}
  \caption{The RG-flows of the scalar and vector yukawa couplings $\bar h_{S,k},\,\bar h_{V,k}$. The thick and thin lines correspond to different initial values of the couplings at different initial scales.}\label{fig:yuk}
\end{center}
\end{figure}

\begin{figure*}[t]
\begin{center}
\subfigure{\includegraphics[width=0.48\textwidth]{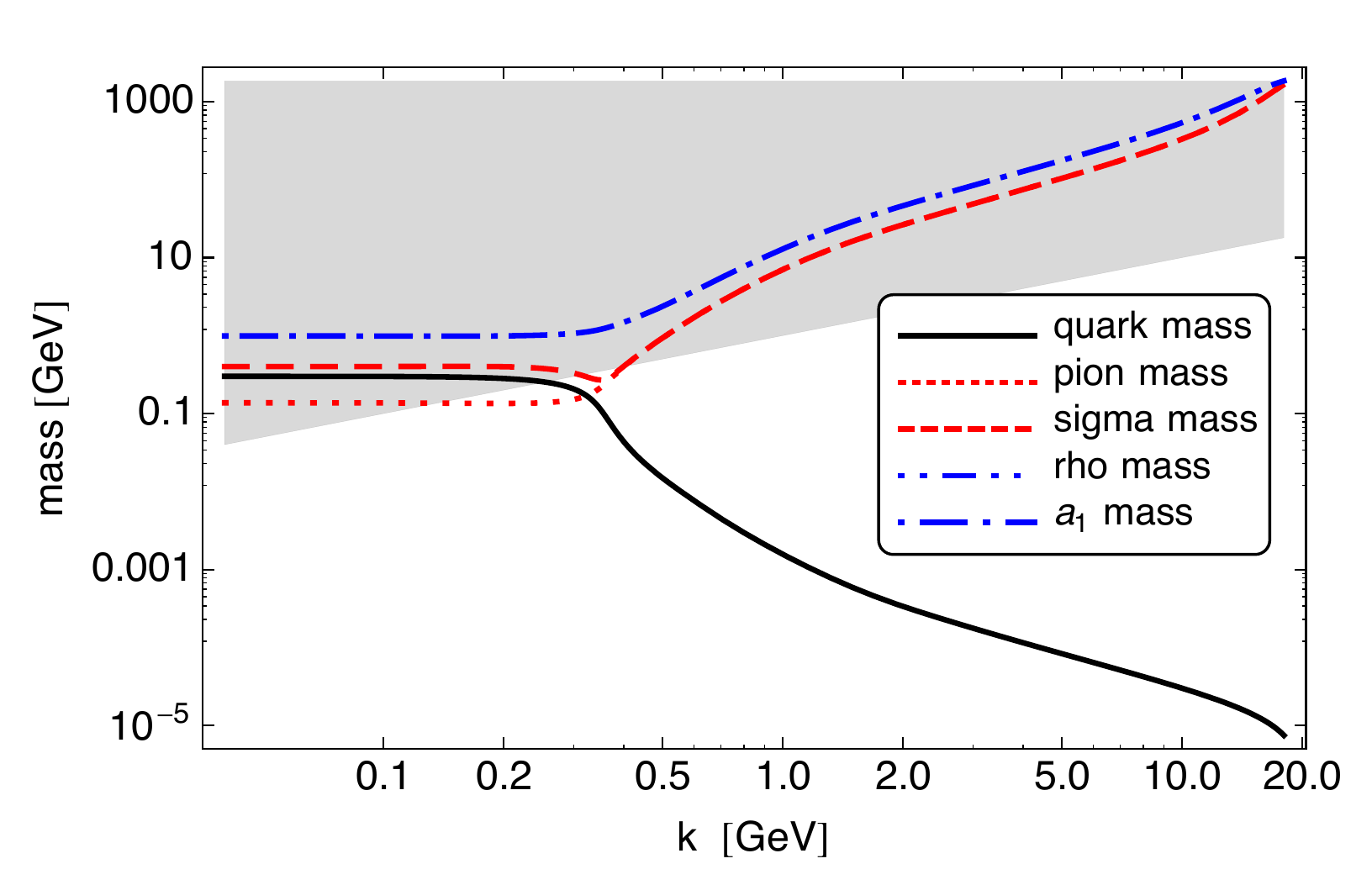}}
\hspace{0.02\textwidth}
\subfigure{\includegraphics[width=0.46\textwidth]{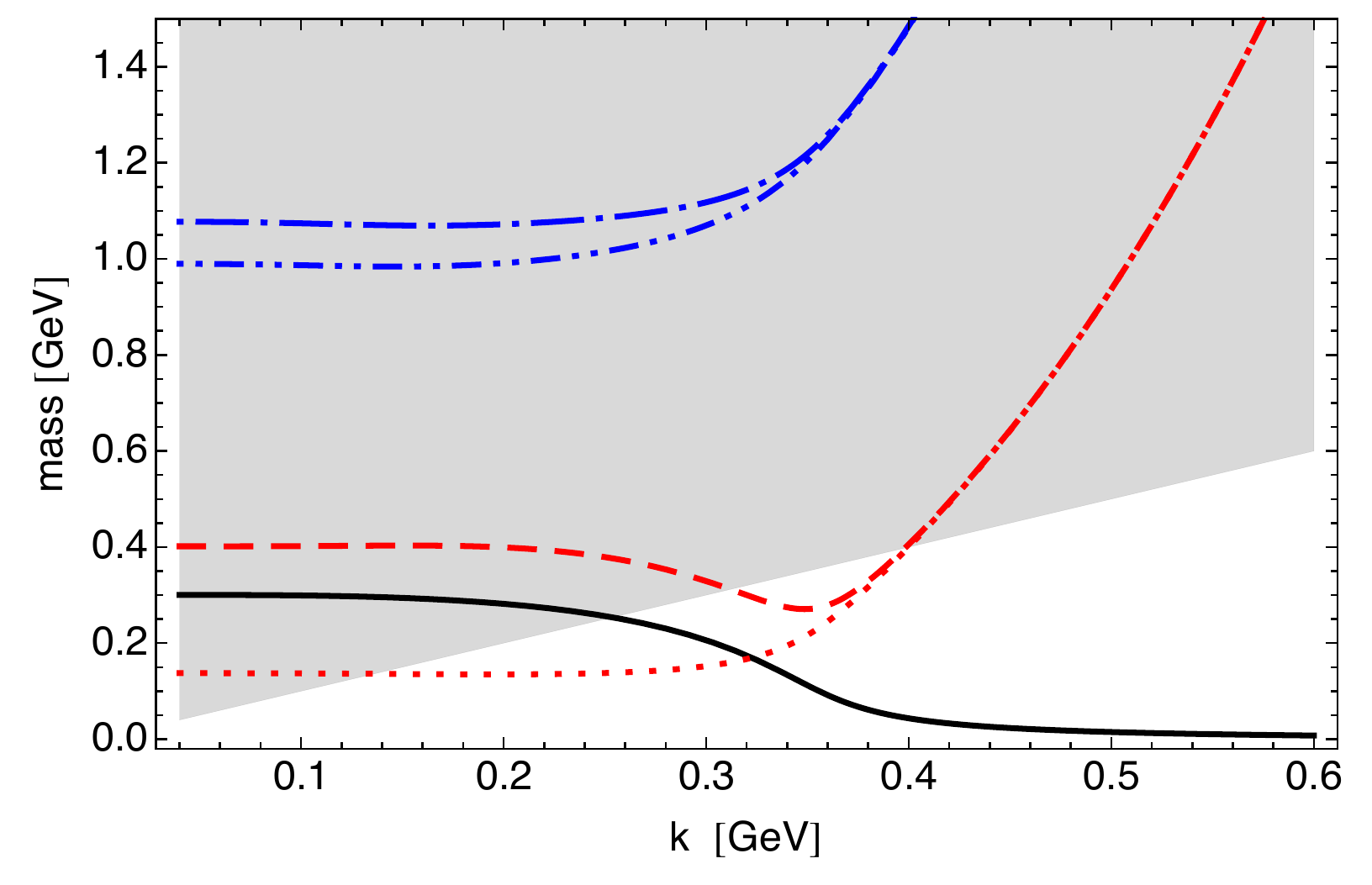}}
\caption{Renormalized masses \eq{eq:massdef} as a function of the RG-scale. Masses in the shaded area are larger than the cutoff scale and therefore decoupled from the dynamics.}\label{fig:masses}
\end{center}
\end{figure*}

\begin{figure}[t]
\begin{center}
  \includegraphics[width=0.98\columnwidth]{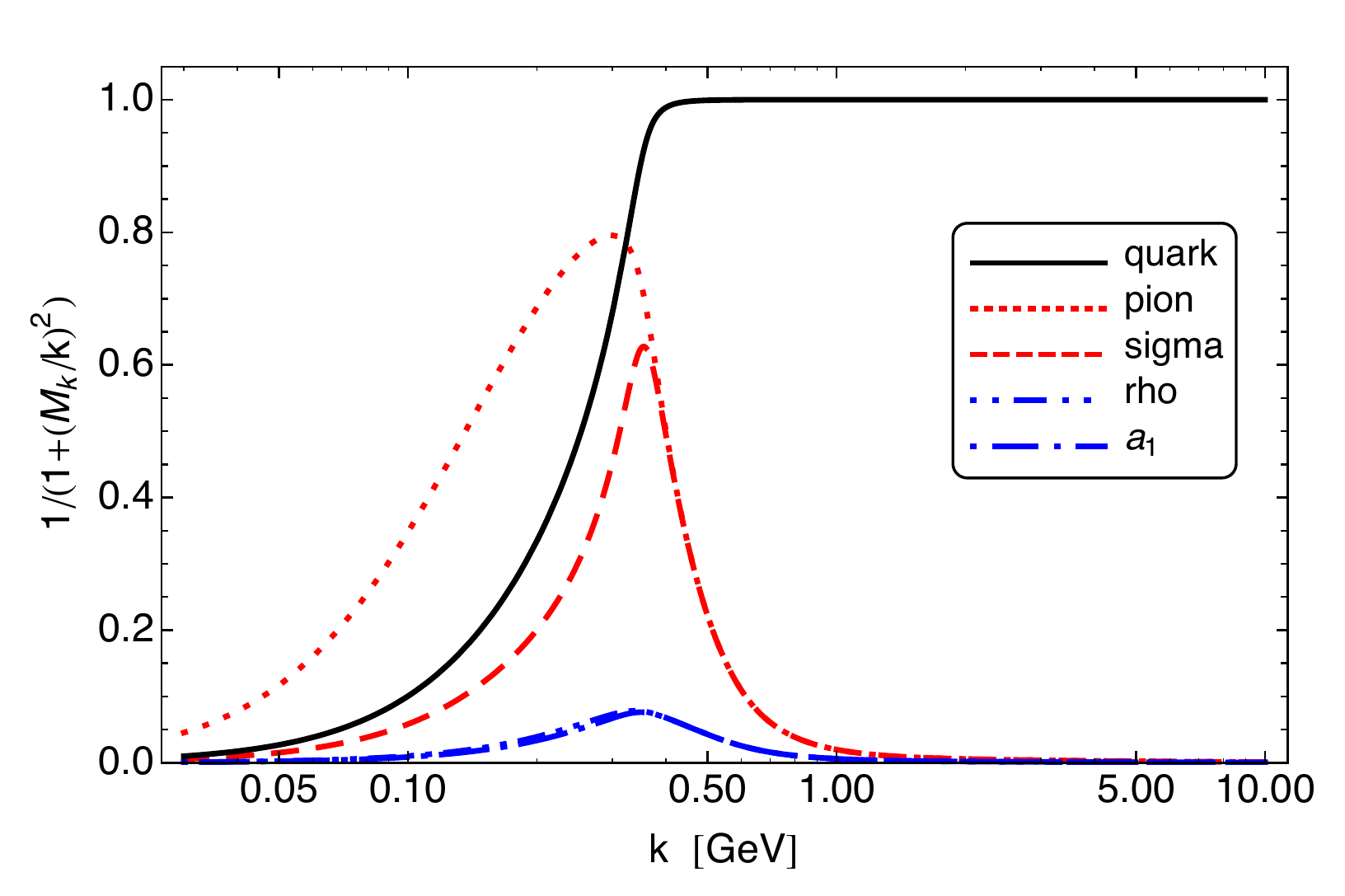}
  \caption{Effective propagators \eq{eq:effprop} of the matter fields as a function of the RG scale. They are a measure for the effective strength of the fluctuations of the fields.}\label{fig:props}
\end{center}
\end{figure}

We initiate the RG flow of the effective action \eq{eq:trunc} at the initial scale $\Lambda\! =\! 20\,\text{GeV}$ and therefore deep in the perturbatively accessible quark-gluon regime. Hadronic degrees of freedom will become relevant at much lower, non-perturbative scales. Owing to the dynamical hadronization procedure, the RG flows of our model are completely fixed by specifying the free parameters of the microscopic gauge fixed action of QCD, i.e. the strong coupling and the current quark mass. Even though we choose a priori different flow equations for the strong couplings $\alpha_{\bar q A q},\, \alpha_{A^3},\, \alpha_{\bar c A c}$, the gauge principle enforces them to be identical in the perturbative regime. The initial value of the strong couplings implicitly sets the scale, and we choose $\alpha_{i,\Lambda}\! =\! 0.163$ for $i\!= \bar q A q, A^3, \bar c A c,$ which corresponds to $\Lambda \approx 20\,\text{GeV}$. Indeed, \Fig{fig:alphas} shows, that the gauge couplings are all identical and follow the perturbative 1-loop running down to $k\! \approx\! 4\,\text{GeV}$, where non-perturbative effects start to induce differences between these couplings.

The current quark mass $m_q^{\text{UV}}$ is related to the explicit symmetry breaking parameter $c$ in \eq{eq:pot} via
\begin{align}
m_q^{\text{UV}} = \frac{h_{S,\Lambda}}{m_{S,\Lambda}^2}\, c.
\end{align}
We choose for the renormalized parameter $\bar c_\Lambda = 3.9\,\text{GeV}^3$, which yields an pion mass in the IR of $M_{\pi,0} = 137.5\, \text{MeV}$.

We note that the physical parameters are rescaled with appropriate powers on the wave function renormalizations to ensure RG invariance, see App.~\ref{sec:rucou} and in particular \eq{eq:rginv}. The physical (or renormalized) quark and meson masses are defined as
\begin{align}\label{eq:massdef}
M_{q,k} = \frac{m_{q,k}}{Z_{q,k}}\,,\quad \text{and}\quad M_{\phi,k} = \frac{m_{\phi,k}}{Z_{\phi,k}^{1/2}}\,.
\end{align}
With slight abuse of terminology, we refer to $m_k$ as bare mass. They are given in \eq{masses}.

The initial conditions of the mesonic parameters can be chosen arbitrarily. In the regime of weak gauge coupling, the flows of these couplings are governed by an infrared-attractive fixed point \cite{Gies:2002hq}. Thus, as long as the initial scale is large enough, we find unique solutions for the meson parameters at low energies. This is given as long as the initial meson masses are chosen larger than the UV-cutoff scale, $M_{\phi,\Lambda} \gtrsim \Lambda$. That way, the mesons do not contribute to the dynamics of the system at high energies. Furthermore, to ensure that our initial conditions correspond to QCD, the ratio $h_{S/V,\Lambda}^2/m_{\pi/\rho,\Lambda}^2$ has to be much smaller than $\Lambda^{-2}$. It corresponds to the four-quark coupling $\lambda_{S/V,\Lambda}$ at the initial scale. A large initial value of the four-quark coupling would describe a gauged Nambu--Jona-Lasinio model with strong coupling, rather than QCD.

The independence of the IR-physics on the initial values of the meson sector is demonstrated in \Fig{fig:yuk} and \Fig{fig:vmd}. There, we have chosen initial values at different initial scales (10 and 20 GeV) that differ by many orders of magnitude and one nicely sees that the initially different trajectories are attracted towards a unique solution in the hadronic regime.

\subsection{Masses}\label{sec:masses}

In \Fig{fig:masses} we show our results for the quark and meson masses. The left figure shows the masses over the full range of scales we consider here, while the right figure shows the region for $k<600\,\text{MeV}$. For scales $k\gtrsim 400\,\text{MeV}$ all mesons are decoupled from the flow. At these scales the dynamics are driven completely by current quarks and gluons. At about $400\,\text{MeV}$, the degeneracy of the $\pi$ and $\sigma$ masses as well as the $\rho$ and $a_1$ masses is lifted due to chiral symmetry breaking. $\pi$, $\sigma$ and the constituent quarks are the dynamical degrees of freedom in this region. The vector mesons are always decoupled. Thus, the vacuum structure of the vector mesons is determined by quark and gluon fluctuations at large scales and the fluctuations of the lightest mesons, the $\pi$ and $\sigma$, at lower scales.

This is also shown in \Fig{fig:props}, where the effective propagators also reflect this scale hierarchy. They are defined in App.~\ref{sec:rucou} and in particular \eq{eq:effprop} and are a measure for the strength of the fluctuations of the fields. Vanishing of the effective propagator of a field implies that this field does not contribute to the dynamics of the system. Thus, we see that at large energy scales the quarks are the only dynamical matter fields. There is only a relatively small window, $100\,\text{MeV}<k<500\,\text{MeV}$, where the scalar mesons are dynamical and one nicely sees that the sigma mesons decouples earlier that the lighter pions. Vector meson fluctuations are always negligible.

Indeed, an explicit calculation of a complete set of four-quark interactions in Euclidean spacetime shows that the scalar-pseudoscalar channel is the dominant channel \cite{Mitter:2014wpa}. This implies that the only relevant meson degrees of freedom in vacuum are $\pi$ and $\sigma$. We note that this picture will change in Minkowski space, since different channels will become relevant as soon as the momentum is close to the corresponding mass pole.

\begin{figure}[t]
\begin{center}
  \includegraphics[width=0.98\columnwidth]{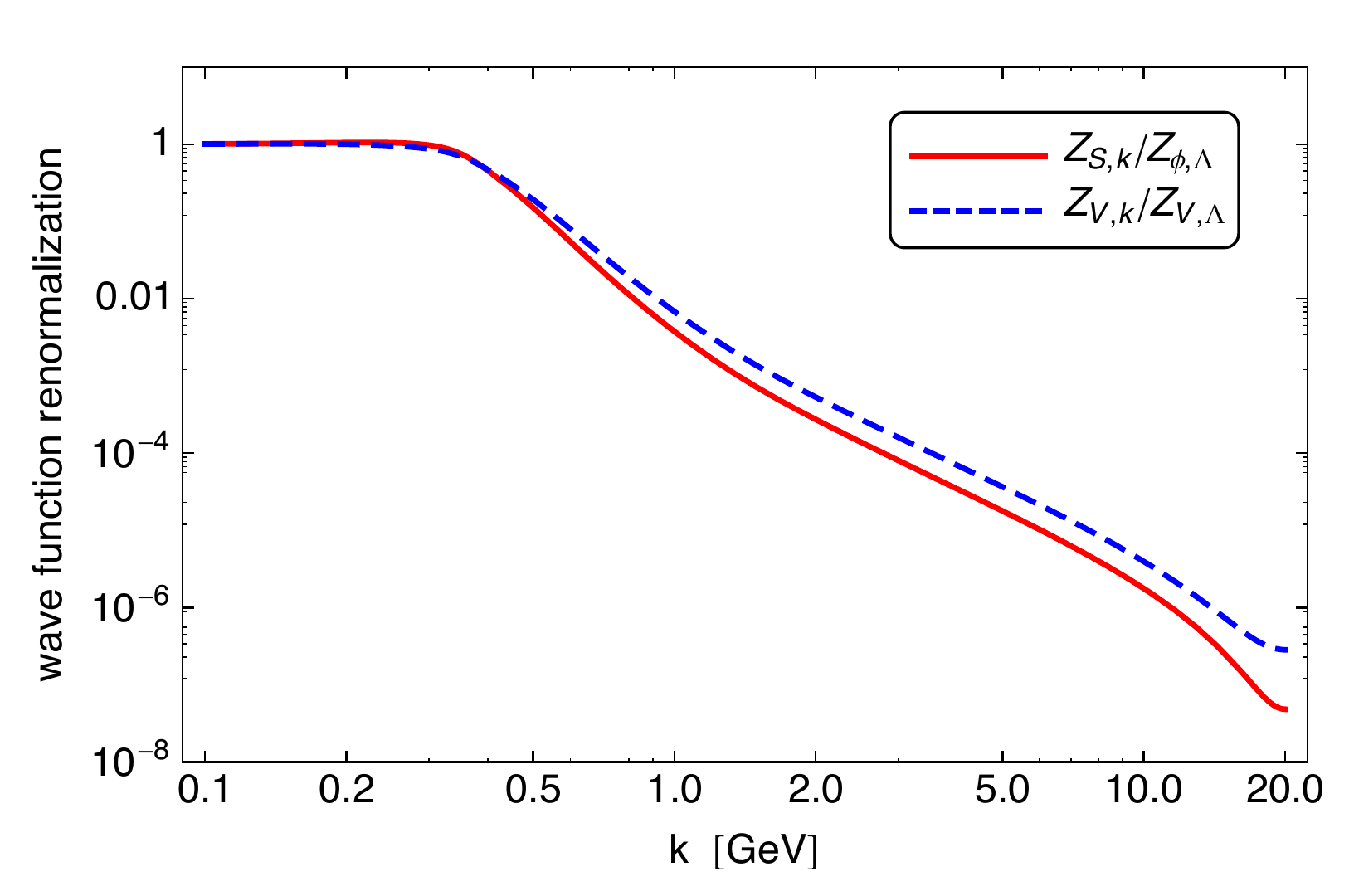}
  \caption{Scalar and vector wave function renormalization as function of the RG scale. We normalized them to be 1 at the IR scale $k\!=\! 30\,\text{MeV}$.}\label{fig:zs}
\end{center}
\end{figure}

The behavior of the masses as function of the RG scale $k$ reflects their scaling with temperature $T$. In particular, the running of the masses at vanishing temperature is qualitatively very similar to the temperature dependence at $k=0$. In \Fig{fig:masses} we see that the $\rho$ mass is almost constant in the hadronic phase and only grows slowly close to the pseudocritical scale. This in line with the in-medium behavior of the $\rho$ mass observed within effective field theory studies, which reproduce the data on vector meson spectral functions and dilepton spectra very well \cite{Rapp:1999ej}. For a sensible comparison, however, we also need to compute the in-medium modifications of the masses within our QCD-based approach.

Our predicted masses for the vector mesons show a quite large discrepancy from the observed masses. We find for the renormalized $\rho$ mass $M_{\rho,0} = 990\,\text{MeV}$, which is about 29\% larger than the observed mass of $770\,\text{MeV}$ \cite{pdg}. For the $a_1$ mass we find $M_{a_1} = 1077\,\text{MeV}$, which is about 15\% smaller than the observed mass of 1260 MeV. The value of the $\rho$ mass is fixed mainly by the fluctuations in the quark-gluon sector. This can be seen from the definition of the masses, \eq{masses}, the observation that $\bar m_{V,k}$ runs only very little in the hadronic regime and that $\bar g_{3,k}$ is very small (see \Fig{fig:vmd}). Thus, the strength of the four-quark interaction $\lambda_{V,k}$, which is determined by the strong coupling, essentially fixes the $\rho$ mass. Furthermore, according to \eq{masses}, the mass-splitting of $\rho$ and $a_1$ and therefore the mass of $a_1$ is determined by the flow of the hadronic sector at low energies. We note that the situation is different for $\pi$ and $\sigma$: the mass of the pion is fixed by its nature as a (pseudo) Goldstone boson and the strength of explicit symmetry breaking in terms of a finite current quark mass. In any case, the mass-splitting of chiral partners in the phase with broken chiral symmetry is sensitive to the quality of our truncation in the hadronic sector. Thus, the small mass of $a_1$ is a signal for a shortcoming of our truncation there and may be related to momentum dependencies that were taken into account insufficiently. The large value for the $\rho$ mass can also be attributed to the insufficient inclusion of momentum dependence, but in the quark-gluon sector and in particular in the four-fermi interaction $\lambda_{V,k}$. We evaluate this interaction at vanishing external momentum, see App.~\ref{sec:rucou}, and it is possible that we underestimate its strength this way. A larger $\lambda_{V,k}$ leads to smaller vector meson masses. Since the present work is the first study in this direction, aimed at capturing the qualitative features, we defer a thorough quantitative analysis to future work. 

\begin{figure}[t]
\begin{center}
  \includegraphics[width=0.98\columnwidth]{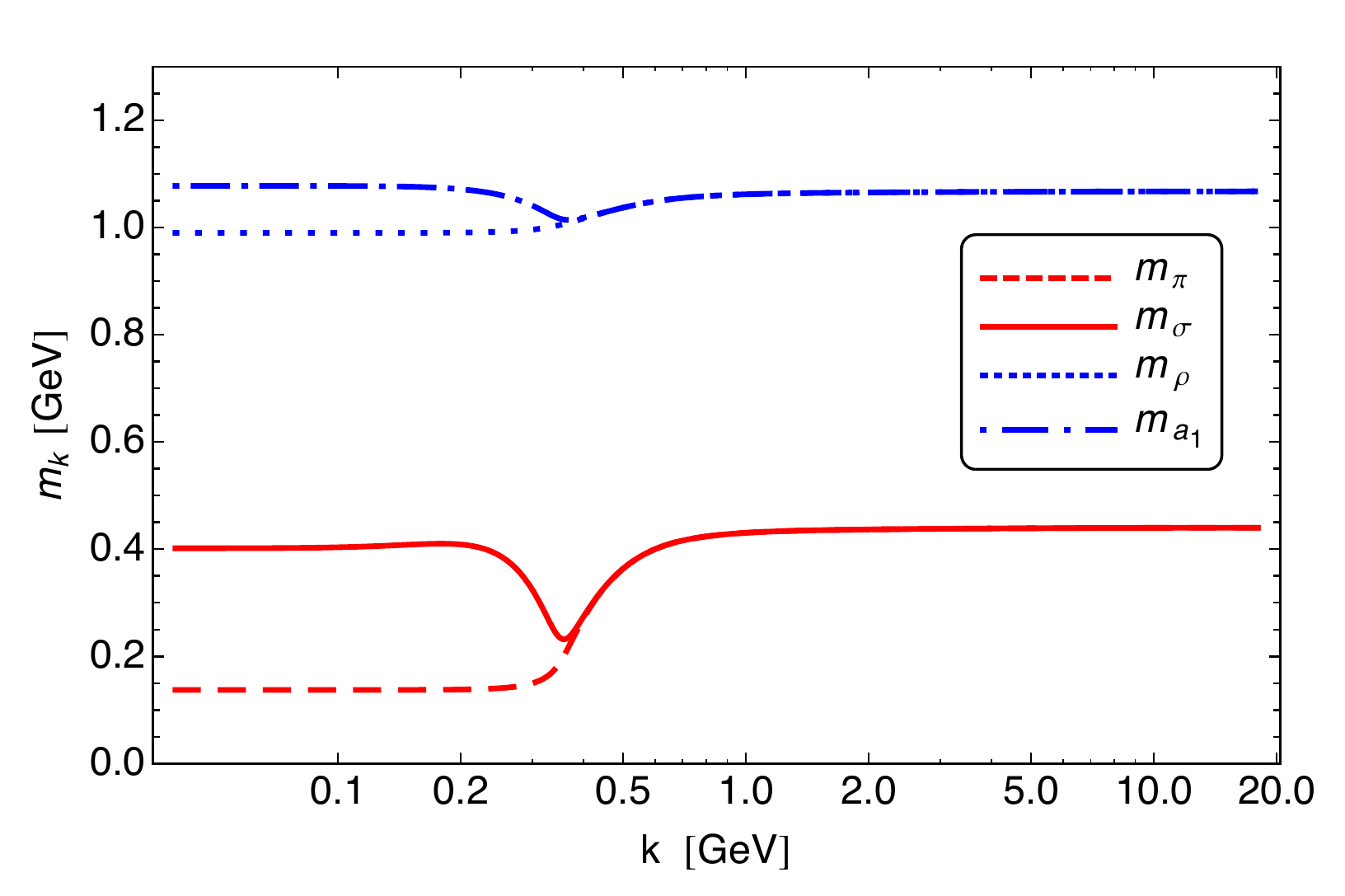}
  \caption{Bare masses of the mesons, $m_{\phi}^2=Z_{\phi,k}\, M_{\phi,k}^2$, see \eq{masses} and \eq{eq:massdef}. Note that we normalized the wave function renormalizations to be 1 in the IR at $k\!=\!20\,\text{MeV}$ here.}\label{fig:baremass}
\end{center}
\end{figure}

\subsection{Decoupling of the Mesons}\label{sec:mesdec}

Mesons are not present in the quark-gluon plasma. In a formulation of the dynamics of QCD on a very wide range of scales in terms of one scale dependent effective action, as in the present work, however, hadronic parameters are necessarily a part of the action also at very large scales. As we have demonstrated here, the meson masses are much larger than the cutoff scale in the quark-gluon regime and therefore the mesons are completely decoupled in this phase. We want to emphasize that this physically desirable picture is achieved with dynamical hadronization. The decoupling of the mesons is triggered by a rapid fall-off of the meson wave function renormalizations $Z_{S/V,k}$ at the pseudocritical scale. Their running is shown in \Fig{fig:zs}. While they stay almost constant in the hadronic regime, they rapidly fall-off at about 400 MeV. The scalar meson wave function renormalization $Z_{S,k}$ drops about eight orders of magnitude and that of the vector mesons, $Z_{V,k}$, about seven orders of magnitude towards the UV. The fastest drop-off is in the vicinity of the pseudocritical scale $k_\chi \approx 400$ MeV. The reason is that quark fluctuations decrease the meson wave function renormalizations in the quark-gluon regime. Their flows are proportional to the corresponding squared Yukawa couplings, $\partial_t Z_{S/V,k} \propto -\bar h_{S/V,k}^2\,Z_{S/V,k}$, at scales $k \gtrsim k_\chi$, resulting in large negative beta functions, see Fig.~\ref{fig:yuk}. Since the wave function renormalizations are the coefficients of the kinetic terms in the effective action \eq{eq:trunc}, their vanishing implies that the mesons become auxiliary fields and are therefore not part of the physical spectrum at large energy scales.

This is reflected in the behavior of the bare masses, i.e. the masses without rescaling with the wave function renormalizations, $m_{\phi}^2=Z_{\phi,k}\, M_{\phi,k}^2$ \eq{eq:massdef}, shown in \Fig{fig:baremass}. The bare masses would not decouple in the quark-gluon regime: while they do not differ from the renormalized masses (\Fig{fig:masses}) in the hadronic regime where the wave function renormalizations are almost constant and of order one, they are constant in the quark-gluon regime. Thus, at large scales the bare meson masses are always much smaller than the the cutoff scale. Without the rapid fall-off of the meson wave function renormalizations, the mesons show no decoupling, resulting in an unphysical high-energy phase. Note that the constant bare masses imply in particular that the running of the physical masses is exclusively driven by the anomalous dimensions of the corresponding mesons at large energy scales.

Since the wave function renormalizations only enter the set of flow equations through the corresponding anomalous dimensions, the flow equations for the wave function renormalizations do not need to be integrated for the solution of the system and all results are independent of the initial values $Z_\Lambda$. For illustration purposes, we have chosen the initial conditions such that $Z_{S/V,0} = 1$.

\subsection{Vector Meson Dominance}\label{sec:vmd}

\begin{figure}[t]
\begin{center}
  \includegraphics[width=0.98\columnwidth]{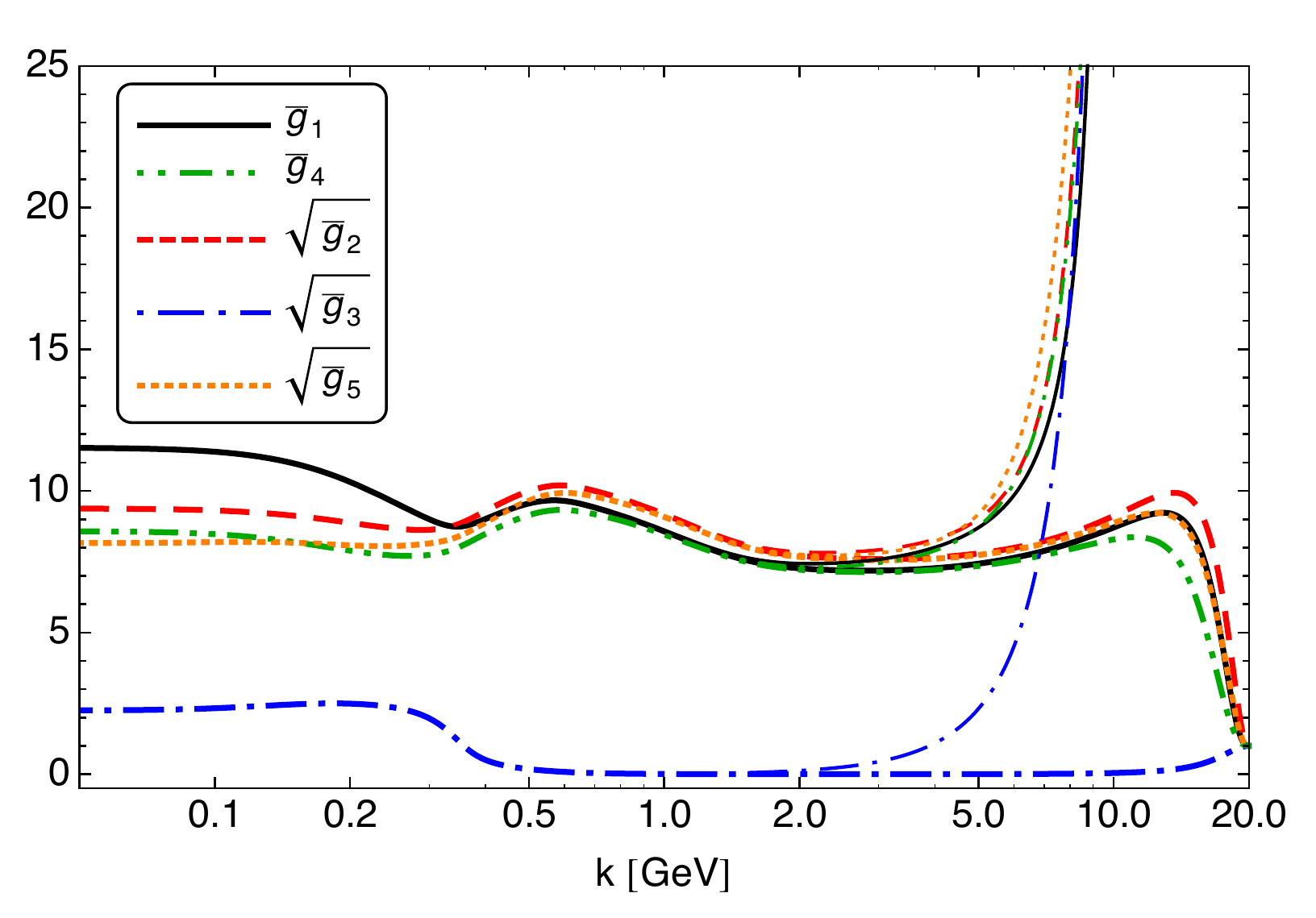}
  \caption{The RG-flow of the vector-vector and vector-scalar meson couplings $\bar g_{1-5,k}$. The thick and thin lines correspond to different initial values of the couplings at different initial scales. Vector meson dominance would imply \eq{eq:vmd2}.}\label{fig:vmd}
\end{center}
\end{figure}

The principle of vector meson dominance (VMD) entails that the $SU(N_f)_A \times SU(N_f)_V$ flavor symmetry is treated as a gauge symmetry. In this case, the vector and axial-vector mesons appear as gauge bosons of the scalar and pseudoscalar mesons. This simplifies the effective action in the hadronic sector, since the gauge principle significantly restricts the number of possible different interactions and there is only one running coupling for interactions involving vector mesons. Here, we do not apply VMD. As a consequence, we have a priori different running couplings $g_{1-5,k}$, while VMD implies
\begin{align}\label{eq:vmd2}
g_{1,k} = g_{4,k} = \sqrt{g_{2,k}}= \sqrt{g_{5,k}}\quad \text{and}\quad g_{3,k}=0\,.
\end{align}
As we have discussed above, the advantage of our approach is that the hadronic parameters are uniquely determined by the dynamics in the quark-gluon phase, i.e. microscopic QCD. Thus, even though we have a large parameter space in the meson sector, model parameter tuning is not necessary. This allows us to study the validity of VMD in an unbiased way by comparing our results to \eq{eq:vmd2}. In \Fig{fig:vmd} we show our results for the running of $\bar g_{1-5,k}$. 

Or results show that, while VMD does not hold exactly, it is a good approximation. In particular the couplings $\bar g_{2,k},\,\bar g_{4,k}$ and $\bar g_{5,k}$ are very close together. Only $\bar g_{1,k}$ is considerably larger than the other couplings. If we define the error one would make by assuming VMD by the standard deviation of these couplings in the IR, we find it to be about 16\% of the mean average of these couplings. $g_{3,k}$, which is explicitly forbidden for local chiral symmetry, is well approximated by VMD. It is very close to $g_{3,k}=0$ at the pseudocritical scale and assumes only a small finite value at lower scales. The flow of $g_{3,k}$ is proportional to the chiral order parameter $\sigma_{0,k}$. Thus, with large positive anomalous dimensions, the renormalized coupling $\bar g_{3,k}$ is driven to values very close to zero at large energy scales.

The construction of our effective action is based on a small momentum expansion (derivative expansion) and we define all running coupling at vanishing external momentum, see App.~\ref{sec:rucou}. The momentum scale of our results is therefore given by $k$. Thus, our findings in the hadronic regime correspond to small momentum scales $k\leq 400\,\text{MeV}$. A comparison of effective field theory predictions assuming VMD with experimental results for the electromagnetic form factor of the pion show that they agree within 10-20\% accuracy at momentum transfer $q^2\lesssim 1\,\text{GeV}^2$ \cite{Meissner:1987ge}. Thus, our results for the validity of VMD are in very good agreement with phenomenological findings.

We note again that the thick and thin lines in \Fig{fig:vmd} correspond to very different initial conditions for the flow of the couplings. The flows in the hadronic phase as well as the final value of the couplings in the IR are prediction of our analysis without any model parameter fixing.

\section{Conclusions and Outlook}\label{sec: concl}

A thorough understanding of the dynamics of vector mesons in QCD is essential for our understanding of the phase structure of strongly interacting matter. Since these low-energy degrees of freedom ultimately derive from microscopic QCD, the dynamical connection between the high- and low-energy sector of QCD needs to be captured. To this end, we have presented the first functional renormalization group study of vector mesons in QCD. Our focus was on how the dynamics of the lightest meson chiral partners, $\pi$, $\sigma$ and $\rho$, $a_1$, emerge from the dynamics of quarks and gluons. We have developed a scale dependent effective action that captures the dynamical transition from the quark-gluon regime to the hadronic regime, including vector mesons, in a qualitative manner. The key ingredient is the dynamical hadronization technique, which allows for a consistent description of the transition from high-energy to low-energy degrees of freedom. This entails in particular that the properties of the hadronic regime are fixed by the quark-gluon fluctuations at high energies. Thus, no fine-tuning of model parameters is necessary and e.g. the meson masses and the running of the mesonic parameters can be viewed as predictions from first-principle QCD.

We have demonstrated explicitly that, within this Euclidean formulation, there is an intriguing scale hierarchy emerging, where the hadronic contributions to the dynamics of the heavier mesons are determined solely by the fluctuations of $\pi$ and $\sigma$.

The masses of $\rho$ and $a_1$ are almost constant and only slightly grow towards the pseudocritical scale. Since the behavior of the masses as a function of the RG-scale reflects their finite temperature behavior, this gives a hint for the in-medium scaling of these masses. Our predictions are in agreement with the findings of phenomenologically motivated effective models.

We have emphasized the important role that the meson wave function renormalizations play for the decoupling of the meson degrees of freedom at high energies. They fall-off many orders of magnitude in the vicinity of the pseudocritical scale. This triggers a rapid growth of the renormalized meson masses and the mesons become auxiliary fields in the quark-gluon phase.

Since the properties of the mesonic parameters in our model are fixed by the QCD flow, we have been able to make an unbiased analysis of the validity of vector-meson dominance. Our results show that while VMD does not hold exactly, it is a good approximation within an accuracy of about 16\% at small momentum scales. This is in agreement with phenomenological findings. 

In this study we focused on qualitative features and given the lack of quantitative precision, in particular for the mass-splitting of the mesons in the chirally broken phase, there is a lot of room for improvement and refinement. In particular the extension of our truncation in the hadronic sector and a thorough analysis of momentum dependencies are important next steps.

This work serves as a starting point for the study of the in-medium modifications of the vector mesons and their spectral functions within functional renormalization group methods for QCD.


\vspace{0.1cm} {\it Acknowledgments -} I am indebted to Robert D. Pisarski who initiated and inspired this work and for helpful comments on the manuscript. Furthermore, I thank Jan M. Pawlowski and Mario Mitter for discussions, comments and collaboration on related projects. I gratefully acknowledge the hospitality of the Brookhaven National Laboratory, where most of this work has been carried out. My work is supported by Helmholtz Alliance HA216/EMMI and the HGS-HIRe Abroad program, which funded the stay at BNL.

\begin{appendix}

\section{Running couplings}\label{sec:rucou}

In this section we provide some details about the RG flow equations of the running couplings of our truncation \eq{eq:trunc} together with the modifications that result from dynamical hadronization \eq{eq:hadflow}. Due to excessive length of the explicit flow equation, we only present their definitions here. For the derivation of most of the equations we used an extension of DoFun \cite{Huber:2011qr} which utilizes Form \cite{2000math.ph..10025V} and FormLink \cite{2012arXiv1212.3522F}. This extension was developed and first used by the authors of \cite{Mitter:2014wpa}. With the truncation \eq{eq:trunc}, the Wetterich equation \eq{eq:fleq} and the definitions given below, the flow equations of the couplings are uniquely specified.

Physical parameters are RG-invariant quantities. To achieve this, all fields are rescaled with their respective wave function renormalizations, $\Phi \rightarrow \sqrt{Z_{\Phi,k}} \Phi$, and all couplings are rescaled with appropriate powers of the wave function renormalizations accordingly, see below. This entails in particular, that the wave function renormalizations enter the flow equations only through the corresponding anomalous dimensions,
\begin{align}\label{eq;etas}
\eta_k = -\frac{\partial_t Z_k}{Z_k}\,.
\end{align}
The physical, i.e. RG-invariant, parameters of the action are defined as
\begin{align}
\bar{\xi}_k = \frac{\xi_k}{\left(Z_{A,k}^{n_A} Z_{q,k}^{n_q} Z_{c,k}^{n_c} Z_{\phi,k}^{n_\phi} Z_{V,k}^{n_V}\right)^{1/2}}\,,
\end{align}
where
\begin{align}\label{eq:rginv}
\begin{split}
\xi_k \in &\, \{ g_{\bar q A q,k},\,g_{A^3,k},\,g_{A^4,k},\,g_{\bar c A c,k},\,\lambda_{S,k},\, \lambda_{V,k},\, h_{S,k},\\
&\quad h_{V,k},\, m_{S,k},\, m_{V,k},\, \nu_{k},\, c_{k},\, g_{1-5,k} \}\,,
\end{split}
\end{align}
is one of the running couplings of our truncation \eq{eq:trunc}. $n_A,\,n_q,\,n_c,\,n_\phi,\,n_V$ are the numbers of gluon, quark, ghost, scalar meson and vector meson fields respectively, that are attached to the coupling $\xi_k$. The physical masses are given by \eq{eq:massdef}. Note that the definition of the gluonic vertices \eq{eq:vertices} implies that the gauge couplings are already RG-invariant. Thus, in that case $\bar{g}_k=g_k$ and we omit the bars.

We use 4d regulator functions of the form
\begin{align}
\begin{split}
R_{k}^A(p^2) &= Z_{A,k}\, p^2 r_B(p^2/k^2)\,\Pi^\perp\,,\\
R_k^q(p^2) &= Z_{q,k}\, \gamma_\mu p_\mu r_F(p^2/k^2)\,,\\
R_{k}^c(p^2) &= Z_{c,k}\, p^2 r_B(p^2/k^2)\,,\\
R_k^\phi(p^2) &= Z_{\phi,k}\, p^2 r_B(p^2/k^2)\,,\\
R_{k}^V(p^2) &= Z_{V,k}\, p^2 r_B(p^2/k^2)\,\Pi^\perp\,,
\end{split}
\end{align}
with the transversal projection operator
\begin{align}
\Pi_{\mu\nu}^\perp = \delta_{\mu\nu}-\frac{p_\mu p_\nu}{p^2}\,.
\end{align}
For the bosonic and fermionic regulator shape functions $r_B$ and $r_F$ we use the optimized shape functions \cite{Litim:2000ci}
\begin{align}\label{eq:litr}
\begin{split}
r_B(x)&=\left(\frac{1}{x}-1\right)\Theta(1-x)\,,\\
r_F(x)&=\left(\frac{1}{\sqrt{x}}-1\right)\Theta(1-x)\,.
\end{split}
\end{align}
The flow equations presented in the following are derived using these specific regulators. They have the advantage, that the loop-momentum integration can be performed analytically and, consequently, all beta functions can be given in analytical form. Furthermore, we work in Landau gauge, fix the Euclidean spacetime dimension to $d\!=\!4$ and color and flavor are fixed to $N_c\!=\!3$ and $N_f\!=\!2$.

First, we explain the effective propagators used in Sec.~\ref{sec:masses} and in particular Fig.~\ref{fig:props}. The propagators in momentum space are of the form
\begin{align}
\begin{split}
G_{B,k}(q) &= \frac{1}{Z_{B,k} q^2 \left(1+r_B(q^2/k^2)\right)+m_{B,k}^2}\,,\\
G_{F,k}(q) &= \frac{1}{Z_{F,k}^2 q^2 \left(1+r_F(q^2/k^2)\right)^2+m_{F,k}^2}\,,
\end{split}
\end{align}
for bosons and fermions respectively. For the specific choice of regulator shape functions \eq{eq:litr}, they read
\begin{align}\label{eq:litprop}
\begin{split}
G_{B,k}(q) &= \frac{\Theta(k^2-q^2)}{Z_{B,k}k^2+m_{B,k}^2}+\frac{\Theta(q^2-k^2)}{Z_{B,k} q^2+m_{B,k}^2}\,,\\
G_{F,k}(q) &= \frac{\Theta(k^2-q^2)}{Z_{F,k}^2k^2+m_{F,k}^2}+\frac{\Theta(q^2-k^2)}{Z_{F,k}^2 q^2+m_{F,k}^2}\,.
\end{split}
\end{align}
To be consistent with the low-momentum expansion our construction of the effective action is based on, we define all running couplings at vanishing external momentum. This entails that all integrands of the loop-momentum integrations are proportional to $\Theta(k^2-q^2)$ which stems from the scale derivative of the regulator $\partial_t R_k^\Phi$ in the flow equation \eq{eq:fleq}. Thus, only the first term of the propagators in \eq{eq:litprop} contributes in the final flow equations. We therefore define the effective propagators relevant for the flows of the physical quantities as
\begin{align}\label{eq:effprop}
\bar G_{\Phi,k}= \frac{1}{1+(M_{\Phi,k}/k)^2}\,.
\end{align} 
Vanishing $\bar G_{\Phi,k}$ implies that the field $\Phi$ does not contribute to the dynamics of the system.

We proceed with the definition of the flows of the gauge couplings. The explicit form of the flow equations is given in \cite{Braun:2014ata}. Here, we only present our definitions for completeness. As we have discussed in Sec.~\ref{sec:gauge}, we compute all three-point functions of QCD, but restrict them to have only the classical tensor structure. We therefore define the flow of the quark-gluon vertex $g_{\bar q A q}$ as
\begin{align}
\partial_t g_{\bar q A q,k} = \frac{1}{8 N_f (N_c^2-1)} \lim_{p\rightarrow 0} \text{Tr}\!\left.\left( \gamma_\mu t^a \frac{\delta^3\partial_t \Gamma_k}{\delta q \delta A_\mu^a \delta\bar q} \right)\right|_{\Phi=\Phi_0},
\end{align}
where $\Phi_0\!=\!(0,0,0,0,0,0,\sigma_{0,k},0,0)$ is the vacuum expectation value of the mean field $\Phi=(A,q,\bar{q},c,\bar c,\pi,\sigma,\rho,a_1)$. The trace runs over all external indices and includes a loop-momentum integration. The limit denotes that all external momenta are set to zero. We define the three-gluon vertex $g_{A^3,k}$ via the projection
\begin{align}
 \partial_t g_{A^3,k}&=\frac{i}{12 N_c (N_c^2-1)}\lim_{p\rightarrow
    0}\frac{\partial^2}{\partial p^2}\\ \nonumber
  &\quad\left.\text{Tr}\left(\delta_{\mu\nu}p_\sigma f^{abc}\frac{
        \delta^3\partial_t\Gamma_{k}}{\delta A(p)_\mu^a\delta
        A(-p)_\nu^b\delta A_\sigma^c(0)}\right)\right|_{\Phi=\Phi_0}\,.
\end{align}
Since as an approximation we evaluate all flow equations at vanishing external momentum, the ghost-gluon vertex $g_{\bar c A c,k}$ only has canonical running. The diagrams that contribute to the beta function are proportional to the external momentum and therefore vanish here and we are left with
\begin{align}
\partial_t g_{\bar c A c} = \frac{1}{2}\left( \eta_{A,k}+2\eta_{C,k} \right)\, g_{\bar c A c}.
\end{align}
Since we approximate the four-gluon vertex with the three-gluon vertex, see \eq{eq:g4a}, we do not need a separate equation for this coupling.

Next, we discuss the flow of the four-quark couplings. Here, we consider two channels, the scalar--pseudoscalar channel with coupling $\lambda_{S,k}$ and the iso-vector--iso-axialvector channel with coupling $\lambda_{V,k}$. Some caution is advised when four-fermion interactions are included in the effective action. A specific quark-antiquark interaction channel can always be expressed as an linear combination of different interaction channels with two spinor fields interchanged. This can potentially lead to ambiguities in the corresponding bosonized models since different sets of composite states can be related to one and the same fermionic action (see e.g. \cite{Braun:2011pp}). This is known as the Fierz ambiguity. While this ambiguity can lead to large uncertainties in mean-field calculations, appropriate approximations that go beyond mean-field in RG studies can minimize these uncertainties \cite{Jaeckel:2003uz}.

Indeed, as explicit calculations considering the RG flows of a Fierz-complete basis of four-quark interactions have shown \cite{Mitter:2014wpa}, the scalar-pseudoscalar $(S\!-\!P)$ channel is the dominant channel in vacuum, while all other channels are strongly suppressed compared to this channel. Furthermore, the dynamical hadronization of only the $(S\!-\!P)$ is sufficient to render all four-quark interaction channels finite at the chiral phase transition. Thus, the error we make from not using a Fierz-complete basis is expected to be small. We therefore restrict our model to contain only two physically relevant channels. In order to study the properties of the corresponding composite fields, we dynamically hadronize both channels here.

We define the running coupling of the scalar-pseudoscalar channel via the projection
\begin{align}\label{eq:lsproj}
 \left.\partial_t\right|_{\phi} \lambda_{S,k}&=\frac{1}{8 N_f N_c (2 N_f N_c+1)}\\ \nonumber
&\quad\times \lim_{p\rightarrow
    0} \left.\text{Tr}\left(\delta_{AB}\delta_{CD}\frac{\delta^4
        \partial_t\Gamma_{k}}{\delta q_A \delta\bar q_B \delta q_C \delta\bar q_D}\right)\right|_{\Phi=\Phi_0}\,,
\end{align}
where $A,\,B,\,C,\,D$ abbreviate the color, flavor and spinor indices of the quarks. For the vector-axialvector channel we choose
\begin{align}\label{eq:lvproj}
 \left.\partial_t\right|_{\phi} \lambda_{V,k}&=-\frac{1}{3} \left.\partial_t\right|_{\phi} \lambda_{S,k}\\ \nonumber
&\quad- \lim_{p\rightarrow
    0} \left.\text{Tr}\left(\mathbb{P}_V^{ABCD} \frac{\delta^4
        \partial_t\Gamma_{k}}{\delta q_A \delta\bar q_B \delta q_C \delta\bar q_D}\right)\right|_{\Phi=\Phi_0}\,,
\end{align}
with the projection operator
\begin{align}
\mathbb{P}_V^{ABCD} =\frac{1}{192 N_f N_c}\, \delta_{AB} \delta_{CD} \gamma_\mu^{\alpha_A \alpha_B} \gamma_\mu^{\alpha_C \alpha_D}\,.
\end{align}
Here, $\alpha_{A,B,C,D}$ is the spinor index of the respective quark field. The Kronecker deltas are summed over the remaining color and flavor indices.

We note that these projections give the flow equations for scale-independent meson fields, i.e. without dynamical hadronization. Dynamical hadronization enforces \eq{eq:hadcond}. Nevertheless, the flows of the four-quark interactions defined in \eq{eq:lsproj} and \eq{eq:lvproj} play a major role for the dynamics of the system and enter the hadronized flow equations in the meson sector via the hadronization functions \eq{eq:hadfunc}.

Following the discussion in \cite{Pawlowski:2014zaa}, we define the scalar Yukawa coupling $h_{S,k}$ via the quark-antiquark two-point function as:
\begin{align}
\partial_t h_{S,k} = \frac{-i}{4 N_f N_c\, \sigma_0} \lim_{p \rightarrow 0} \text{Tr}\!\left.\left(\!\delta_{AB} \frac{\delta^2\partial_t \Gamma_k}{\delta q_A \delta\bar q_B} \right)\right|_{\Phi=\Phi_0}\,.
\end{align}
Taking dynamical hadronization into account, the total flow of the renormalized scalar Yukawa coupling is
\begin{align}\label{eq:modhs}
\partial_t\bar h_{S,k} = \left.\partial_t\right|_{\phi} \bar h_{S,k} - k^{-2} M_{\pi,k}^2\, \dot{\bar A}_k\,,
\end{align}
where $\dot{\bar A}_k = k^2 Z_{S,k}^{1/2} Z_{q,k}^{-1} \dot A_k$ and $\dot A_k$ is given by \eq{eq:hadfunc}. According to \eq{masses}, the scalar channel Yukawa coupling defines the quark mass. 

We define the vector Yukawa coupling $h_{V,k}$ via the $\rho q \bar q$ three-point function as
\begin{align}
\partial_t h_{V,k} = \frac{1}{16 N_c (N_f^2-1)} \lim_{p \rightarrow 0} \text{Tr}\!\left.\left(\! \gamma_\mu \vec{\tau}\,\frac{\delta^3\partial_t\Gamma_k}{\delta \vec{\rho}^{\,\mu} \delta q \delta \bar q} \right)\right|_{\Phi=\Phi_0}\,,
\end{align}
where contractions over the remaining indices with Kronecker deltas is implied. Splitting the flow into the contributions with and without dynamical hadronization, we find
\begin{align}\label{eq:modhv}
\partial_t\bar h_{V,k} = \left.\partial_t\right|_{\phi} \bar h_{V,k} - k^{-2} M_{\rho,k}^2\, \dot{\bar B}_k\,,
\end{align}
with $\dot{\bar B}_k = k^2 Z_{V,k}^{1/2} Z_{q,k}^{-1} \dot B_k$ and $\dot B_k$ given by \eq{eq:hadfunc}.

We want to emphasize that the modifications of the Yukawa couplings in \eq{eq:modhs} and \eq{eq:modhs} proportional to $\dot A_k$ and $\dot B_k$ are crucial for the dynamical hadronization procedure. They guarantee that the ratio $h_{S/V,k}^2/m_{\pi/\rho,k}^2$ replaces the four-quark interactions $\lambda_{S/V,k}$, which vanish due to dynamical hadronization, in the quark-gluon phase. This way, the modified Yukawa couplings capture the relevant quark-gluon dynamics at large energy scales, while they act as the usual Yukawa couplings in the hadronic regime.

Next, we discuss the mesonic couplings of our truncation. They are not modified by dynamical hadronization. We define the running of the chiral order parameter $\sigma_{0,k}$ via the pion two-point function as
\begin{align}
\begin{split}
\partial_t \sigma_{0,k} &= -\biggl( \nu_k \sigma_{0,k}+\frac{c_k}{\sigma_{0,k}^2} \biggr)^{-1}\\
&\quad\times \frac{1}{N_f^2-1} \lim_{p \rightarrow 0} \text{Tr}\!\left.\left(\!\delta_{ij}\frac{\delta^2\partial_t\Gamma_k}{\delta \pi_i \delta \pi_j} \right)\right|_{\Phi=\Phi_0}\,,
\end{split}
\end{align} 
with the adjoint flavor indices $i,j$. The flow of the scalar four-point function $\nu_k$ is defined as follows:
\begin{align}\label{eq:nu}
\partial_t \nu_k = \frac{1}{N_f^4-1} \lim_{p \rightarrow 0} \text{Tr}\!\left.\left(\!\delta_{ij}\delta_{kl}\frac{\delta^4\partial_t\Gamma_k}{\delta \pi_i \delta \pi_j \delta \pi_k \delta \pi_l} \right)\right|_{\Phi=\Phi_0}\,,
\end{align}
with the adjoint flavor indices $i,j,k,l$.

The explicit symmetry breaking term $c$ is a source term and therefore drops out of the flow equation. The RG-invariant coupling $\bar c_k$ therefore only runs canonically,
\begin{align}
\partial_t \bar c_k = \frac{1}{2} \eta_{S,k}\, \bar c_k \,.
\end{align}

The meson masses are defined as the momentum independent part of the corresponding two-point functions. For the scalar mesons, we need the flow of $m_{S,k}$ which is given by
\begin{align}
\partial_t m_{S,k}^2 = \frac{1}{N_f^2-1} \lim_{p \rightarrow 0} \text{Tr}\!\left.\left(\!\delta_{ij} \frac{\delta^2\partial_t \Gamma_k}{\delta \pi_i \delta\pi_j} \right)\right|_{\Phi=\Phi_0}\,.
\end{align}
We cannot define the flow of the vector meson mass parameter $m_{V,k}$ independently of other couplings, since we have to either project on the $\rho$ or the $a_1$ mass, which gives contributions from other couplings in the chirally broken phase according to \eq{masses}. We choose to project on the $\rho$ mass and find
\begin{align}
\begin{split}
\partial_t m_{V,k}^2 &= \frac{1}{4(N_f^2-1)} \lim_{p \rightarrow 0} \text{Tr}\!\left.\left(\!\delta_{\mu\nu}\delta_{ij} \frac{\delta^2\partial_t \Gamma_k}{\delta \rho_i^\mu \delta\rho_j^\mu} \right)\right|_{\Phi=\Phi_0}\\
&\quad-\sigma_{0,k}^2 \partial_t g_{3,k}\,.
\end{split}
\end{align}
The flow $\partial_t g_{3,k}$ is defined below in \eq{eq:g23dot}.

For the definition of the three-point function $g_{1,k}$ we choose the $\rho\pi\pi$ vertex and find:
\begin{align}\label{eq:g1def}
\nonumber \partial_t g_{1,k} &= \frac{-i}{2 N_f (N_f^2-1)} \lim_{p \rightarrow 0} \frac{\partial^2}{\partial p^2} \\\nonumber
&\quad \text{Tr}\!\left.\left(\!p_\mu \epsilon_{ijk}\frac{\delta^3\partial_t\Gamma_k}{\delta \rho_i^\mu(-p) \delta \pi_j(p) \delta \pi_k(0)} \right)\right|_{\Phi=\Phi_0}\\
&\quad + \sigma_{0,k}^2 \partial_t \left( \frac{g_{1,k} g_{2,k}}{m_{a_1,k}^2} \right)-g_{2,k}\sigma_{0,k}\dot{C}_k\,.
\end{align}
As we have discussed in Sec.~\ref{sec:pia1}, the elimination of the $\pi\!-\!a_1$ mixing leads to two types of modifications of the $\rho\pi\pi$ vertex. The first stems from the modifications of the action due to the replacement \eq{eq:pia1rep} and leads to a modification of this vertex given by \eq{eq:rppm}. The first term in the third line of \eq{eq:g1def} cancels the additional term in the flow to ensure that we compute the flow of $g_{1,k}$ and not of \eq{eq:rppm}. The second modification stems from the scale dependence of $a_1$ that is introduced by \eq{eq:pia1rep}. This leads to the second term in the third line of \eq{eq:g1def} which follows from \eq{eq:hadflow}.

We define the couplings $g_{2,k}$ and $g_{3,k}$ via the flow
\begin{align}\label{eq:g23dot}
\partial_t g_{2/3,k} =  \lim_{p \rightarrow 0}\text{Tr}\!\left.\left(\! \delta_{\mu\nu} \mathbb{P}_{g_{2/3}}^{ijkl}\frac{\delta^4 \partial_t \dot \Gamma_k}{\delta \pi_i \delta \pi_j \delta \rho_k^\mu \delta \rho_k^\nu} \right)\right|_{\Phi=\Phi_0}\,,
\end{align}
with the projection operator for $g_{2,k}$
\begin{align}
\mathbb{P}_{g_{2}}^{ijkl} = \frac{1}{4 N_f (N_f^2+1)} \left( \frac{1}{N_f^2-1} \delta_{ij} \delta_{kl}-\delta_{ik}\delta_{jl} \right)\,,
\end{align}
and the projection operator for $g_{3,k}$
\begin{align}
\mathbb{P}_{g_{3}}^{ijkl} = \frac{1}{4 (N_f^4+1)} \left( \frac{1}{2} \delta_{ij} \delta_{kl}+\delta_{ik}\delta_{jl} \right)\,.
\end{align}

The vector meson self-interactions $g_{4,k}$ and $g_{5,k}$ are defined as
\begin{align}
\partial_t g_{4,k} &=  \frac{-i}{6 N_f (N_f^2-1)} \lim_{p \rightarrow 0} \frac{\partial^2}{\partial p^2}\\ \nonumber
&\quad\times \text{Tr}\!\left.\left(\! p_\alpha \delta_{\beta\gamma} \epsilon_{ijk} \frac{\delta^3 \partial_t \dot \Gamma_k}{ \delta \rho_i^\alpha(p)  \delta \rho_j^\beta(-p)  \delta \rho_k^\gamma(0)} \right)\right|_{\Phi=\Phi_0}\,,
\end{align}
and
\begin{align}
\partial_t g_{5,k} &=  \frac{1}{24 N_f (N_f^2-1)} \lim_{p \rightarrow 0}\\ \nonumber
&\quad\times \text{Tr}\!\left.\left(\! \delta_{\alpha\beta}\delta_{\gamma\delta} \delta_{ij} \delta_{kl} \frac{\delta^4 \partial_t \dot \Gamma_k}{ \delta \rho_i^\alpha \delta \rho_j^\beta \delta \rho_k^\gamma \delta \rho_l^\delta} \right)\right|_{\Phi=\Phi_0}\,.
\end{align}

Finally, we discuss the wave function renormalizations. As mentioned before, in a RG-invariant formulation they enter the flow equations only via the corresponding anomalous dimensions \eq{eq;etas}. The ghost anomalous dimension and the gauge part of the gluon anomalous dimension are discussed in Sec.~\ref{sec:gauge}. the quark contribution to the gluon anomalous dimension $\Delta \eta_{A,k}$, i.e. the vacuum polarization, is computed from
\begin{align}\label{fig:deltaeta}
\includegraphics[height=8.3ex]{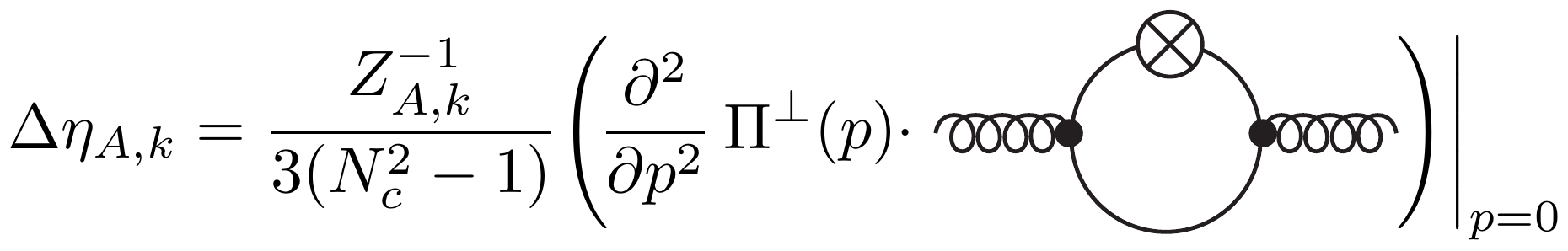}
\end{align}
Again, the lines correspond to the full propagators and the dots to the full vertices. The explicit equation can be found in \cite{Braun:2014ata}.

The quark anomalous dimension is computed from
\begin{align}
\eta_{q,k} &= \frac{-1}{8 N_f N_c Z_{q,k}}\\ \nonumber
&\quad\times \lim_{p \rightarrow 0}\frac{\partial^2}{\partial\! p^2}\text{Tr}\!\left.\left(\! \gamma_\mu p_\mu\frac{\delta^2 \partial_t \dot \Gamma_k}{\delta \bar q(p) \delta q(-p)} \right)\right|_{\Phi=\Phi_0}\,,
\end{align}
where contraction of external color, flavor, and spinor indices is understood.

The scalar meson anomalous dimension $\eta_{S,k}$ has to be defined via the pion-pion two-point function. Using the sigma meson two-point function leads to additional contributions to the flow with couplings that correspond to a higher order derivative expansion. We define the anomalous dimension for scale-independent fields as
\begin{align}\label{eq:etas}
\left.\eta_{S,k}\right|_\phi &= \frac{-1}{2 (N_f^2-1) Z_{S,k}}\\ \nonumber
&\quad\times \lim_{p \rightarrow 0}\frac{\partial^2}{\partial\! p^2}\text{Tr}\!\left.\left(\! \delta_{ij} \frac{\delta^2 \partial_t \dot \Gamma_k}{\delta \pi_i(p) \delta \pi_j(-p)} \right)\right|_{\Phi=\Phi_0}\,.
\end{align}
It receives modifications from the scale dependence of $a_1$ from the elimination of the $\pi\!-\!a_1$ mixing. The full anomalous dimension then is
\begin{align}
\eta_{S,k} = \left.\eta_{S,k}\right|_\phi - \bar g_{2,k}\bar\sigma_{0,k}\dot{\bar C}_k\,
\end{align}
where the second term follows from \eq{eq:hadflow}. We use the rho meson to define the vector meson anomalous dimension and find

\begin{align}
\eta_{V,k} &= \frac{-1}{6 (N_f^2-1) Z_{V,k}}\\ \nonumber
&\quad\times \lim_{p \rightarrow 0}\frac{\partial^2}{\partial\! p^2}\text{Tr}\!\left.\left(\! \delta_{ij}\delta^{\mu\nu} \frac{\delta^2 \partial_t \dot \Gamma_k}{\delta \rho_i^\mu(p) \delta \rho_j^\nu(-p)} \right)\right|_{\Phi=\Phi_0}\,.
\end{align}

We emphasize that due to chiral symmetry the definition of the mesonic couplings in terms of $n$-point functions is not unique. We have explicitly checked that different projection procedures give the same results as long as they are equivalent by chiral symmetry. However, some caution is advised since seemingly equivalent definitions may give different results. The reason in those cases is that that inappropriate projections may contaminate the flows with contributions that are not part of the truncation. For example, a definition of $\nu_k$ via the sigma meson four-point function instead of \eq{eq:nu} gives additional contributions from diagrams that are related to the flow of the 6-meson interaction. Another example is $\eta_{S,k}$, which is mentioned above \eq{eq:etas}. To find appropriate projection procedures one therefore has to keep extended truncations, such as general field-dependent couplings, in mind.

\end{appendix}


\bibliography{vector2}

\end{document}